\documentclass[draft,tightenlines,nofootinbib,preprint,aps,eqsecnum,amsmath,amssymb]{revtex4}

\newcommand{\beq}{\begin{equation}}
\newcommand{\eeq}{\end{equation}}
\newcommand{\bea}{\begin{eqnarray}}
\newcommand{\eea}{\end{eqnarray}}

\begin{document}

\title{
The Physical Role of Gravitational and Gauge Degrees of Freedom in
General Relativity - I: Dynamical Synchronization and Generalized
Inertial Effects.}

\medskip

\author{Luca Lusanna}

\affiliation{ Sezione INFN di Firenze\\ Polo Scientifico\\ Via Sansone 1\\
50019 Sesto Fiorentino (FI), Italy\\ E-mail LUSANNA@FI.INFN.IT}

\author{Massimo Pauri}

\affiliation{
Dipartimento di Fisica - Sezione Teorica\\ Universita' di Parma\\
Parco Area Scienze 7/A\\ 43100 Parma, Italy \\E-mail
PAURI@PR.INFN.IT}

\begin{abstract}

This is the first of a couple of papers in which the peculiar
capabilities of the Hamiltonian approach to general relativity are
exploited to get both {\it new} results concerning specific
technical issues, and new insights about {\it old} foundational
problems of the theory. The first paper includes:

1) a critical analysis of the various concepts of symmetry related
to the Einstein-Hilbert Lagrangian viewpoint on the one hand, and
to the Hamiltonian viewpoint, on the other. This analysis leads,
in particular, to a re-interpretation of {\it active}
diffeomorphisms as {\it passive and metric-dependent} dynamical
symmetries of Einstein's equations, a re-interpretation which
enables to disclose the (up to now unknown) connection of a
subgroup of them to Hamiltonian gauge transformations {\it
on-shell};

2) a re-visitation of the canonical reduction of the ADM
formulation of general relativity, with particular emphasis on the
geometro-dynamical effects of the gauge-fixing procedure, which
amounts to the definition of a \emph{global (non-inertial)
space-time laboratory}. This analysis discloses the peculiar
\emph{dynamical nature} that the traditional definition of distant
simultaneity and clock-synchronization assume in general
relativity, as well as the {\it gauge relatedness} of the
"conventions" which generalize the classical Einstein's
convention.

3) a clarification of the physical role of Dirac and gauge
variables, as their being related to {\it tidal-like} and {\it
generalized inertial effects}, respectively. This clarification is
mainly due to the fact that, unlike the standard formulations of
the equivalence principle, the Hamiltonian formalism allows to
define a generalized notion of "force" in general relativity in a
natural way;

\today

\end{abstract}

\maketitle

\vfill\eject

\section{Introduction.}

This is the first of a couple of papers in which we aim to show
the peculiar capability of the Hamiltonian ADM formulation of
metric gravity to grasp a series of conceptual and technical
problems that appear to have not been directly discussed so far.
Some of such problems, although possibly not of primary importance
for the working relativist, are deeply rooted into the
foundational level of the theory and seems particularly worth of
clarification in connection with the alternative programs os
string theory and loop quantum gravity. Some other problems are in
fact \emph{new} problems which give rise to interesting new
solutions about general issues. Our two papers should be read in
sequence, since the first contains various technical premises for
the second. One of the main foundational issues we want to revisit
in the second paper (hereafter referred to as II) is the
well-known {\it Hole Argument} or {\it Lochbetrachtung}, raised by
Einstein in 1915-1916 \cite{1a} and, after two years of struggle,
dismissed by him mainly on pragmatic grounds. The deep conceptual
content of the argument has been rebirth by a seminal paper by
Stachel \cite{2a}, and essentially seized since then by the
philosophers of science. On the other hand, in the physical
literature, the {\it Hole Argument} has been bypassed by the
recognition that a 4-geometry does not correspond to a single
tensor solution of Einstein's equation but rather to a whole
equivalence class of solutions in a definite sense (see, e.g.,
\cite{3a}). We believe, however, that the problem deserves further
investigation even from a physicist's point of view and in paper
II we shall show indeed that there is still some beef to bite
around the issue.

Previous partial accounts of the material of this and the
following paper can be found in Refs. \cite{4a,5a,6a}.

The first reason we have to adopt the Hamiltonian approach to
general relativity is that all of the problems we are interested
in are deeply entangled with the initial value problem of the
theory. On the other hand, we do believe that the constrained ADM
methodology is just the {\it only} proper way to analyze all the
relevant aspects of such a problem. This is no surprise, after
all, and it is not by chance that the modern treatment of the
initial value problem within the Lagrangian configurational
approach \cite{7a} must in fact mimic the Hamiltonian methods (see
more in Section II).

Second, in the context of the Hamiltonian formalism, we can
exploit the nearly unknown Bergmann and Komar's theory of {\it
general coordinate group symmetries} \cite{8a}. This help us in
clarifying the various concepts of symmetry related to the
Einstein-Hilbert Lagrangian viewpoint, on the one hand, and to the
Hamiltonian view point, on the other. In particular, it enables us
to show that {\it active diffeomorphisms}, as dynamical symmetries
of Einstein's equations, admit a subgroup which can be interpreted
in a passive way as the Legendre pull-back of Hamiltonian gauge
transformations {\it on shell}. This is the first relevant outcome
that will also turn out to be a crucial premise for the
discussions given in II.

Third, only in the Hamiltonian approach can we isolate the {\it
gauge variables}, which carry the descriptive arbitrariness of the
theory, from the {\it Dirac observables} (DO), which are gauge
invariant quantities carrying the intrinsic degrees of freedom of
the gravitational field, and are subjected to hyperbolic (and
therefore "{\it determinate}", or "{\it causal}" in the customary
sense) evolution equations. The superiority of the Hamiltonian
approach is essentially due to the fact that it allows working
{\it off shell}, i.e., avoiding immediate transition to the space
of solutions of Einstein's equations.

All of our results are obtained by working within a class of
space-times of the Christodoulou-Klainermann type \cite{9a}, which
are globally hyperbolic space-times asymptotically flat at spatial
infinity, enjoying some other interesting properties. Such
space-times can be foliated in Cauchy 3-hypersurfaces
$\Sigma_{\tau}$ (where $\tau$ plays the role of {\it parameter
time}) which play also the role of {\it simultaneity} surfaces and
are the basic starting point of the ADM canonical formulation.
These surfaces are mathematically described by an embedding
$x^{\mu} = z^{\mu}(\tau ,\vec \sigma )$ ($\vec \sigma$ arbitrary
3-coordinates adapted to the $\Sigma_{\tau}$ surfaces). Once the
embedding is given, one can evaluate the unit normals and the
extrinsic curvature of $\Sigma_{\tau}$, and two specific {\it
congruences of time-like observers}. The first, defined by the
field of unit normals, is a {\it surface-forming congruence}; the
second, defined by the field of $\tau$-gradients of the embedding
functions, is in general a {\it rotating congruence}, viz. a
non-surface-forming one. Starting from this mathematical
background, the ADM formulation is realized by a multilevel
circular procedure which, bringing to the solution of the
Einstein-Hamilton equations in terms of 4 initial data for the DO
on a given $\Sigma_{\tau_0}$, backfires to a {\it dynamical
identification} of the initial chrono-geometrical 3+1 setting.
\medskip

The procedure starts with the Hamiltonian transcription of
Einstein's equations in terms of 20 canonical variables, functions
of the components of the 4-metric and their derivatives and
adapted to the 3+1 splitting. Note, incidentally, that unlike such
canonical variables, the initial embedding functions $x^{\mu} =
z^{\mu}(\tau ,\vec \sigma )$ stay as {\it external} elements of
the game until the canonical procedure reaches its aim with the
solution of Einstein's equations.

Since the original Einstein's equations are not hyperbolic, it
turns out that the canonical variables are not all functionally
independent, but satisfy eight {\it constraints}, given as
functions of the canonical variables that vanish on a
12-dimensional {\it constraint surface} (not a phase space!) to
which the physically meaningful states are restricted. When used
as generators of canonical transformations, the eight constraints
map points on the constraint surface to points on the same
surface; these transformations are known as \emph{Hamiltonian
gauge transformations}. If, following Dirac, we make the
reasonable demand that the evolution of all \emph{physical
variables} be unique, then - barring subtler complications - the
points of the constraint surface lying on the same \emph{gauge
orbit}, i.e. linked by gauge transformations, \emph{must describe
the same physical state}. Conversely, only the functions in phase
space that are invariant with respect to gauge transformations can
describe physical quantities. To eliminate this ambiguity and
create a one-to-one mapping between points in the phase space and
physical states, further constraints must be imposed, known as
\emph{gauge conditions} or \emph{gauge-fixings}. The number of
independent gauge-fixing must be equal to the number of
independent constraints, i.e. 8. Such gauge-fixings can be
implemented by arbitrary functions of the canonical variables,
except that they must define a 4-dimensional \emph{reduced phase
space} that intersects each gauge orbit exactly once (\emph{orbit
conditions}) and is coordinatized by the above mentioned Dirac
observables (DO). Technically, this coordinatization is carried
through by the so-called Shanmugadhasan transformation\cite{10a}
which (though almost implicitly) ends with the construction of a
new array of 20 canonical variables in which the 4 canonically
conjugate DO are separated from the eight (Abelianized)
constraints and their conjugated variables\footnote{As a matter of
fact, things are subtler: see Section IIC}. These latter are
precisely the eight {\it gauge variables} that parametrize the
gauge orbits generated by the constraints. The {\it gauge-fixing}
of the {\it gauge variables}, together with the enforcement of the
eight constraints, {\it reduce} the 20-dimensional phase space to
the 4-dimensional phase space of the intrinsic degrees of freedom
of the theory.

The analysis of the canonical reduction and of the
geometro-dynamical meaning of the gauge fixings is instrumental to
clarify an important issue. Actually, a {\it complete gauge
fixing}, has the following implications: i) it removes all the
gauge arbitrariness of the theory by determining the {\it
functional form} of all the gauge variables in terms of the Dirac
observables, which, at this stage of the procedure, are 4
arbitrary fields of $(\tau ,\vec \sigma )$; ii) given the
geometro-dynamical meaning of the gauge variables and of their
variations (see Section IID), the determination of their {\it
functional form} in terms of DO entails the {\it implicit}
fixation of all the elements characterizing the 3+1 splitting of
space-time, in particular: a) the form and 4-dimensional packing
of the Cauchy surfaces, together with a standard of (mathematical)
local proper time; b) the choice of the 3-coordinates on the
Cauchy surfaces; c) the determination of the two congruences of
time-like observers; and d) {\it on-shell} (i.e., on the solutions
of Einstein's equations) the unique fixation of a 4-coordinate
system.

In physical terms this set of choices amount eventually to
individuate a {\it network of intertwined and synchronized local
laboratories made up with test matter} (obviously up to a coherent
choice of chrono-geometric standards). We shall call such network
a {\it global (non-inertial) space-time laboratory}. This
interpretation shows that, unlike in ordinary gauge theories where
the gauge variables are inessential degrees of freedom, the
concept of reduced phase space is very abstract and not directly
useful in general relativity: it is nothing else than the space of
gravitational equivalence classes each of which is described by
the set of all laboratory networks living in a gauge orbit.

These effects of the gauge-fixing procedure entail in turn a
physically interesting consequence which typically characterizes
the canonical description of metric gravity. Actually, once the
{\it complete gauge fixing} has determined the {\it functional
form} of the gauge variables in terms of DO, we are eventually
left with the problem of solving the Einstein equations for the DO
themselves, in terms of their initial values, on some Cauchy
surface $\Sigma_{\tau_0}$. It is only this fundamental step that
brings to its end the whole ADM construction, for the solution
determines in particular the extrinsic curvature of the surfaces
$\Sigma_{\tau}$, which, in its turn, can make explicit the
embedding functions $x^{\mu} = z^{\mu}(\tau ,\vec \sigma )$. This
fixes, as it were, {\it explicitly} the space-time {\it universe}
corresponding to the given initial values of the DO, including the
definition of {\it simultaneity}, distant clock synchronization
and gravito-magnetism.

It is important  to stress, therefore, that the {\it complete
determination of the chrono-geometry} depends upon the solution of
Einstein-Hamilton equations of motion i.e., once the Hamiltonian
formalism is fixed by the gauge choices, {\it upon the initial
conditions for the Dirac observables}. This implies that the
admissible notions of distant simultaneity turn out to be {\it
dynamically determined}. Every solution of Einstein equations with
a given set of admissible initial data admits as many dynamical
simultaneity notions as admissible on-shell 3+1 splittings of
space-time. On-shell, each such splitting defines the
synchronization of clocks in the family of complete Hamiltonian
gauges differing only in the choice of the 3-coordinates on the
simultaneity leaves and in the implied choice of the shift
functions (namely in the gravito-magnetic properties) as shown in
Subsection IIID. These dynamically determined simultaneity notions
are much less in number than those admissible in special
relativity, where such notions are non-dynamical due to the
absolute chrono-geometrical structure of Minkowski space-time. The
upshot, however, is  that, in canonical metric gravity in analogy
to what happens in a non-dynamical way within the framework of
parametrized Minkowski theories (see Ref.\cite{11a,12a} and
Appendix A of Ref.\cite{13a}), different admissible {\it
conventions} about distant simultaneity within the {\it same
universe} are merely {\it gauge-related conventions},
corresponding to different {\it complete gauge options}. We
believe that this result throws an interesting new light even on
the old - and outdated - debate about the so-called {\it
conventionality} of distant simultaneity in special relativity,
showing the trading of {\it conventionality} with {\it gauge
freedom}. It is clear that this trading owes its consistency to
the complete Hamiltonian gauge mechanism based on the 3+1
splitting of space-time. Of course, it rests to be shown how the
above dynamical determination can be {\it enforced in practice} to
synchronize actual clocks, i.e., essentially, how to generalize to
the gravity case the formal structure of Einstein-Reichenbach's
convention. This discussion is given in all details in
Ref.\cite{12a} for the case of special relativity and can easily
be extended to general relativity.

Finally, the separation carried out by the Shanmugadhasan
transformation (conjoined with the circumstance that the
Hamiltonian point of view brings naturally to a re-reading of
geometrical features in terms of the traditional concept of {\it
force}), leads to a third result of our investigation which,
again, would be extremely difficult to characterize within the
{\it Lagrangian viewpoint} at the level of the Hilbert action or
Einstein's equations. This result, concerning the overall physical
role of gravitational and gauge degrees of freedom, is something
that should be added to the traditional wisdom of the equivalence
principle asserting the local impossibility of distinguishing
gravitational from inertial effects. Actually, the isolation of
the gauge arbitrariness from the true intrinsic degrees of freedom
of the gravitational field is instrumental to understand and
visualize which aspects of the local effects, showing themselves,
e.g., on test matter, have a {\it genuine gravitational origin}
and which aspects depend solely upon the choice of the reference
frame and could therefore even be named {\it inertial} in analogy
with their non-relativistic Newtonian counterparts. Indeed, two
main differences characterize the issue of {\it inertial effects}
in general relativity with respect to the non-relativistic
situation: the existence of {\it autonomous degrees of freedom} of
the gravitational field independently of the presence of matter
sources, on the one hand, and the {\it local nature of any single
general-relativistic reference system}, on the other. We shall
show that, although the very definition of {\it inertial forces}
(and of {\it gravitational force} in general) appears to be rather
arbitrary in general relativity, it seems natural to characterize
first of all as genuine gravitational effects those which are
directly correlated to the DO, while the gauge variables appear to
be correlated to the general relativistic counterparts of
Newtonian inertial effects. Another aspect of the Hamiltonian
connection "{\it gauge variables - inertial effects}" is related
to the 3+1 splitting of space-time mentioned above. Since a
variation of the gauge variables modifies the foliation and
thereby the identification of the {\it global (non-inertial)
space-time laboratory}, a variation of gauge variables also
modifies the {\it generalized inertial effects} that manifest
themselves locally.

The only weakness of the analysis leading to the physical
characterization of {\it tidal-like} and {\it generalized inertial
effects} is that the separation of the two autonomous degrees of
freedom of the gravitational field from the gauge variables is, as
yet, a gauge (i.e. coordinate) -dependent concept. The known
examples of pairs of conjugate DO are neither invariant under {\it
passive diffeomorphisms} (PDIQ, i.e., coordinate-independent) nor
tensors. In view of clarifying this point, in paper II we will
discuss the relation between the notion of DO and that of the
so-called {\it Bergmann observables} (BO)\cite{14a} which are
defined (although rather ambiguously) to be, again, as uniquely
{\it predictable} from the initial data, but also invariant under
standard {\it passive diffeomorphisms} (PDIQ).

A possible starting point to attack the problem of the connection
of DO with BO seems to be a Hamiltonian re-formulation of the
Newman-Penrose formalism \cite{15a} (that contains only PDIQ)
employing Hamiltonian null-tetrads carried by the surface-forming
congruence of time-like observers. In view of this program, in
paper II we will argue in favor of a {\it main conjecture}
according to which special Darboux bases for canonical gravity
should exist in which the inertial effects (gauge variables) are
described by PDIQ while the autonomous degrees of freedom (DO) are
{\it also} BO. The hoped for validity of this conjecture, besides
amounting in particular to state the internal consistency of
Bergmann's {\it multiple definition} (which is not fully evident
as it stands), would render our distinction about {\it generalized
inertial} and {\it tidal-like} effects an invariant statement,
giving a remarkable contribution to the old-standing debate about
the equivalence principle. Note in addition that, since the
Newman-Penrose PDIQ are tetradic quantities, the validity of the
conjecture would eliminate the existing difference between the
observables for the gravitational field and the observables for
matter, usually built by means of tetrads associated to some
time-like observer. Furthermore, this would also provide a
starting point for defining a metrology in general relativity in a
generally covariant way\footnote{Recall that this is the main
conceptual difference from the non-dynamical metrology of special
relativity}, replacing the empirical metrology \cite{16a} used
till now. Finally, it would also enable to replace the {\it test
matter} of the axiomatic approach to measurement theory (see
Appendix A of paper II ) with dynamical matter.

\bigskip

The plan of the paper is the following. In Section II the
Einstein-Hilbert Lagrangian viewpoint and the related local
symmetries are summarized. Particular emphasis is given to the
analysis of the most general group $Q$ of dynamical symmetries of
Einstein's equations (Bergmann-Komar group), and the {\it passive
view} of {\it active} diffeomorphisms is clarified. Finally, some
remarks are given about the issue of the choice of coordinate
systems and its relation to the \emph{Lagrangian gauge fixings}.
The ADM Hamiltonian viewpoint and its related canonical local
symmetries are synthetically expounded in Section III. Building on
the acquired knowledge about the structure of $Q$, particular
emphasis is given to a discussion of the general Hamiltonian gauge
group and to the correspondence between {\it active
diffeomorphisms} and {\it on-shell gauge transformations}.
Furthermore, the analysis of the chrono-geometrical meaning of a
complete gauge fixing and the particularities of the closure of
the ADM construction are related to the issue of the {\it
dynamical nature} of the {\it conventions} about distant
simultaneity and gravito-magnetism in general relativity. As shown
in Section IV, the results obtained in Section III about the
canonical reduction lead naturally to the physical interpretation
of the DO and the gauge variables as characterizing {\it
tidal-like} and {\it inertial-like} effects, respectively. The up
to now gauge-dependent status of this distinction is stressed at
the end, as well as the possibility of further clarification of
the issue to be discussed in paper II, together all concluding
remarks. Finally, Appendix A contains a miscellanea of properties
of the accelerated observers, extracted from various scattered
sources.

\vfill\eject

\section{The Einstein-Hilbert Lagrangian Viewpoint and the Related
 Local Symmetries.}

The basic assertion of the {\it general covariance} of general
relativity amounts to the statement that Einstein's equations have
a {\it tensor} character. This is a {\it statement of symmetry}
with many facets.

\bigskip

A) {\it Local Noether Symmetries of the Einstein-Hilbert action.}
\medskip

Given a pseudo-Riemannian 4-dimensional manifold $M^4$ with its
maximal coordinate atlas, the Einstein-Hilbert action for {\it
pure gravity without matter}

\beq
 S_H = \int d^4x\, {\cal L}(x) = \int d^4x \, \sqrt{{}^4g}\, {}^4R,
 \label{II1}
 \eeq

\noindent defines a variational principle for the metric 2-tensor
over $M^4$ whose components, in the coordinate chart $x^{\mu}$,
are ${}^4g_{\mu\nu}(x)$ . The associated Euler-Lagrange equations
are Einstein's equations

\beq
 {}^4G_{\mu\nu}(x) {\buildrel {def}\over =} {}^4R_{\mu\nu}(x) - {1\over 2}\, {}^4R(x)\,
 {}^4g_{\mu\nu}(x)  = 0.
 \label{II2}
 \eeq

As well known, the action (\ref{II1}) is invariant under general
coordinate transformations (the {\it passive diffeomorphisms}
${}_PDiff\, M^4$ ), which are a subset of {\it local Noether
symmetries} (second Noether theorem) of the action. This has the
consequence that:
\bigskip

i) Einstein's equations are form invariant under general
coordinate transformations;

ii) the Lagrangian density ${\cal L}(x)$ is singular, namely its
Hessian matrix has vanishing determinant.

\bigskip This in turn entails that:
\bigskip

i) four of the ten Einstein equations are {\it Lagrangian
constraints}, namely restrictions on the Cauchy data;

ii) four combinations of Einstein's equations and their gradients
vanish identically ({\it contracted Bianchi identities}).
\bigskip

In conclusion, there are only two {\it dynamical} second-order
equations depending on the {\it accelerations} of the metric
tensor. As a consequence, the ten components ${}^4g_{\mu\nu}(x)$
of the metric tensor are functionals of two "{\it deterministic}"
dynamical degrees of freedom and eight further degrees of freedom
which are left completely undetermined by Einstein's equations
{\it even once the Lagrangian constraints are satisfied}. This
state of affairs makes the treatment of both the Cauchy problem of
the non-hyperbolic system of Einstein's equations and the
definition of observables  within the Lagrangian context \cite{7a}
extremely complicated.

\bigskip

In modern terminology, general covariance is interpreted as the
statement that {\it a physical solution of Einstein's equations}
properly corresponds to a {\it 4-geometry}, namely the equivalence
class of all the 4-metric tensors, solutions of the equations,
written in all possible 4-coordinate systems. This equivalence
class is usually represented by the quotient ${}^4Geom = {}^4Riem
/ {}_PDiff\, M^4$, where ${}^4Riem$ denotes the space of metric
tensors solutions of Einstein's equations. Then, any two {\it
inequivalent} Einstein space-times are different 4-geometries.

\bigskip

B) {\it Invariance of Einstein's Equations under active
diffeomorphisms.}
\medskip

Let us recall the basic underlying mathematical concept of {\it
active diffeomorphism} and its consequent action on the tensor
fields defined on the differentiable manifold $M^4$ (see for
instance Ref.\cite{3a}). Consider a (geometrical or {\it active})
diffeomorphism $D_A$ which maps points of $M^4$ to points of
$M^4$: $D_A: p {\rightarrow \hspace{.2cm}}\ p' = D_A \cdot p$, and
its tangent map $D_A^{*}$ which maps tensor fields $T {\rightarrow
\hspace{.2cm}} D_A{*} \cdot T$ in such a way that $[T](p)
{\rightarrow \hspace{.2cm}} [D_A^{*} \cdot T](p) \equiv
[T^{'}](p)$. Then $[D_A^{*} \cdot T](p) = [T](D_A^{-1}\cdot p)$.
It is seen that the transformed tensor field $D_A^{*} \cdot T$ is
a {\it new} tensor field whose components in general will have at
$p$ values that are {\it different} from those of the components
of $T$. On the other hand, the components of $D_A^* \cdot T$ have
at $p'$ - by construction - the same values that the components of
the original tensor field $T$ have at $p$: $T^{'}(D_A \cdot p) =
T(p)$ or $T'(p) = T(D_A^{-1}\cdot p)$. The new tensor field $D_A^*
\cdot T$ is called the {\it drag-along} of $T$. For later use it
is convenient to recall that there is another, non-geometrical -
so-called {\it dual} - way of looking at the active
diffeomorphisms. This {\it duality} is based on the circumstance
that in each region of $M^4$ covered by two or more charts there
is a one-to-one correspondence between an {\it active}
diffeomorpshism and a specific coordinate transformation. The
coordinate transformation ${\cal T}_{D_A}: x(p) {\rightarrow
\hspace{.2cm}}\ x'(p) = [{\cal T}_{D_A}x](p)$ which is {\it dual}
to the active diffeomorphism $D_A$ is defined such that $[{\cal
T}_{D_A}x](D_A \cdot p) = x(p)$. In its essence, this {\it
duality} transfers the functional dependence of the new tensor
field in the new coordinate system to the old system of
coordinates. By analogy, the coordinates of the new system $[x']$
are said to have been {\it dragged-along} with the {\it active}
diffeomorphism $D_A$. It is important to note here, however, that
the above {\it dual view} of active diffeomorphisms, as particular
{\it coordinate}-transformations, is defined only {\it implicitly}
(see more below).

\bigskip

In abstract coordinate-independent language, Einstein's equations
(\ref{II2}) can be written as $G = 0$, where $G$  is the Einstein
2-tensor ($G = G_{\mu\nu}(x)\, dx^{\mu} \bigotimes dx^{\nu}$ in
the coordinate chart $x^{\mu}$). Under an {\it active
diffeomorphism} $D_A: M^4 \mapsto M^4$, $D_A \in {}_ADiff\, M^4$,
we get $G = 0 \mapsto D^*_A\, G = 0$ ($D^*_A\, G$ is the
drag-along or push-forward of $G$), which shows that {\it active
diffeomorphisms are symmetries of the tensor Einstein's
equations}. \footnote{Note that a subset of {\it active
diffeomorphisms} are the {\it conformal isometries}, i.e. those
conformal transformations which are also active diffeomorphisms,
namely ${}^4\tilde g = \Omega^2\, {}^4g \equiv \phi^*\, {}^4g$ for
some $\phi \in {}_ADiff\, M^4$ with $\Omega$ strictly positive.
Since the Hilbert action is not invariant under the conformal
transformations which are not {\it ordinary isometries} (i.e.
conformal isometries with $\Omega = 1$ for which ${\cal L}_X\,
{}^4g = 0$, if $X$ is the associated Killing vector field), only
these latter are Noether dynamical symmetries.}.

\bigskip

C) {\it Dynamical symmetries of Einstein's partial differential
equations (PDE)}
\medskip

Einstein's equations, considered as a set of partial differential
equations in a given coordinate chart, conjoined with a choice of
a function space for the solutions, have their own {\it passive
dynamical symmetries} \cite{17a} which only partially overlap with
the local Noether symmetries. Let us stress that:

i) A dynamical symmetry is defined only on the space of solutions
of the equations of motion, namely it is an {\it on-shell}
concept. As a consequence, the very definition of dynamical
symmetries entails the study of the integrability of the equations
of motion. In particular, in the case of  completely Liouville -
integrable systems dynamical symmetries are re-interpretable as
maps of the space of Cauchy data onto itself. Let us stress that
in gauge theories, and especially in Einstein's theory, {\it the
space of Cauchy data is partitioned in gauge-equivalent classes of
data}: all of the Cauchy data in a given class identify a single
Einstein space-time (or 4-geometry). The dynamical symmetries of
Einstein's equations follow therefore in two classes: a) those
mapping inequivalent Einstein space-times among themselves, and b)
those acting within a single Einstein space-time mapping
gauge-equivalent Cauchy data among themselves (actually, they are
{\it on shell} gauge transformations).

ii) Only a subset of such symmetries (called {\it Noether
dynamical symmetries}) can be extended {\it off-shell} in the
variational treatment of the action principle.
\medskip

The {\it passive diffeomorphisms} ${}_PDiff\, M^4$ are just an
instantiation of Noether dynamical symmetries of Einstein's
equations.

\bigskip

Let us observe that in the physical literature on field theory one
is mainly concerned with the natural Noether symmetries of the
Hilbert action, i.e. with {\it passive diffeomorphisms}. On the
other hand, according to Stachel \cite{2a}, it is just the
dynamical symmetry nature of {\it active diffeomorphisms} that
expresses the real {\it physically relevant} content of {\it
general covariance}. This dualism active-passive has been a
continuous source of confusion and ambiguity in the literature,
which we would like to clarify presently.

\bigskip

Let us look preliminarily at some implications of points A) and
B). Choose a reference coordinate chart $x^{\mu}$, where the
metric components are ${}^4g_{\mu\nu}(x)$. Every passive
diffeomorphism defines a new system of coordinates $x^{\mu}
\mapsto x^{'\, \mu} = f^{\mu}(x)$ [with inverse $x^{'\, \mu}
\mapsto x^{\mu} = h^{\mu}(x^{'})$] where the new form of the
metric components is given by the standard tensorial
transformation rule

\beq
 {}^4g^{'}_{\mu\nu}(x^{'}(x)) = {{\partial h^{\alpha}(x^{'})}\over {\partial
 x^{'\, \mu}}}\, {{\partial h^{\beta}(x^{'})}\over {\partial x^{'\, \nu}}}\,
 {}^4g_{\alpha\beta}(x).
 \label{II3}
 \eeq

On the contrary, an active diffeomorphism $D_A\, p \mapsto p^{'}$
defines both a coordinate transformation (the {\it drag-along
coordinate system}) $x^{\mu} \mapsto y_A^{\mu}(x)$ with
$y_A^{\mu}{|}_{p^{'}} = x^{\mu}{|}_p$ and the drag-along $D^*_A\,
{}^4g $ of the metric tensor, whose components are defined through
the equation $dy^{\mu}_A\, dy_A^{\nu}\, (D^*_A\,
{}^4g)_{\mu\nu}(y_A) {|}_{p^{'}} = dx^{\mu}\, dx^{\nu}\,
{}^4g_{\mu\nu}(x){|}_p$. As a consequence, we have: i) the tensor
components $(D^*_A\, {}^4g)_{\mu\nu}(y_A)$ are {\it not} the
components of the metric tensor in the chart $y_A^{\mu}$ implied
by Eq.(\ref{II3}); ii) in the original coordinate chart $(D^*_A\,
{}^4g)_{\mu\nu}(x) \not= {}^4g_{\mu\nu}(x)$.
\bigskip

The hints for a clarification of the active/passive ambiguity can
be found in a nearly forgotten paper by Bergmann and Komar
\cite{8a} [see, however, Ref.\cite{18a}] in which it is shown that
the biggest group $Q$ of {\it passive dynamical symmetries} of
Einstein's equations is not ${}_PDiff\, M^4$ [$x^{{'}\, \mu} =
f^{\mu}(x^{\nu})$] but instead a larger group of transformations
of the form

\bea
 Q:&& x^{{'}\, \mu} = f^{\mu}(x^{\nu}, {}^4g_{\alpha\beta}(x)),
 \nonumber \\
 &&{}\nonumber \\
  {}^4g^{'}_{\mu\nu}(x^{'}(x)) &=& {{\partial h^{\alpha}(x^{'},
  {}^4g^{'}(x^{'}))}\over {\partial
 x^{'\, \mu}}}\, {{\partial h^{\beta}(x^{'}, {}^4g^{'}(x^{'}))}\over
 {\partial x^{'\, \nu}}}\, {}^4g_{\alpha\beta}(x).
 \label{II4}
 \eea

\noindent It is clear that in this way we allow for metric
dependent coordinate systems, whose associated 4-metrics are in
general different from those obtainable from a given 4-metric
solution of Einstein's equations by {\it passive diffeomorphisms}:
actually, the transformations (\ref{II4}) map points to points,
but associate with a given point $x$ an image point $x'$ that
depends also on the metric field \footnote{Strictly speaking,
Eqs.(\ref{II4}) should be defined as transformations on the tensor
bundle over $M4$.}. It is remarkable, however, that not only these
new transformed 4-metric tensors are still solutions of Einstein's
equations, but that, at least for the subset $Q' \subset Q$ which
corresponds to mappings among gauge-equivalent Cauchy data, they
belong indeed to the {\it same 4-geometry}, i.e. the same
equivalence class generated by applying all {\it passive
diffeomorphisms} to the original 4-metrics: $ {}^4Geom  = {}^4Riem
/ {}_PDiff\, M^4 = {}^4Riem / Q'$. Note, incidentally, that this
circumstance is mathematically possible only because ${}_PDiff\,
M^4$ is a {\it non-normal} sub-group of $Q$. The 4-metrics built
by using passive diffeomorphisms are, as it were, only a dense
sub-set of the metrics obtainable by means of the group Q. The
restricted set of active diffeomorphisms passively reinterpreted
with Eq.(\ref{II4}) belongs to the set of local Noether symmetries
of the Einstein-Hilbert action.

\medskip
There is no clear statement in the literature about the dynamical
symmetry status of the group ${}_ADiff\, M^4$ of {\it active
diffeomorphisms} and their relationship with the group $Q$, a
point which is fundamental for our program. To clarify this point,
let us consider an infinitesimal transformation of the type
(\ref{II4}) connecting a 4-coordinate system $[x^{\mu}]$ to a new
one $[x^{{'}\mu}]$ by means of metric-dependent infinitesimal
descriptors:

\beq
 x^{{'}\, \mu} = x^{\mu} + \delta\, x^{\mu} = x^{\mu} + \xi^{\mu}(x, {}^4g).
 \label{II5}
 \eeq

\noindent   This will induce the usual formal variation of the
metric tensor \footnote{What is relevant here is the {\it local}
variation $\bar \delta \, {}^4g_{\mu\nu}(x) = {\cal
L}_{-\xi^{\gamma}\, \partial_{\gamma}}\, {}^4g_{\mu\nu}(x) =
{}^4g^{'}_{\mu\nu}(x) - {}^4g_{\mu\nu}(x)$ which differs from the
{\it total} variation by a {\it convective} term: $\delta\,
{}^4g_{\mu\nu}(x) = {}^4g^{'}_{\mu\nu}(x^{'}) - {}^4g_{\mu\nu}(x)
= \bar \delta\, {}^4g_{\mu\nu}(x) + \delta\, x^{\gamma}\,
\partial_{\gamma}\, {}^4g_{\mu\nu}(x)$.}

\beq
 \bar \delta\, {}^4g_{\mu\nu} = -\Big( \xi_{\mu ;\nu}(x, {}^4g) +
\xi_{\nu ;\mu}(x, {}^4g) \Big).
 \label{II6}
  \eeq

\noindent If $\bar \delta \, {}^4g_{\mu\nu}(x)$ is now identified
with the local variation of the metric tensor induced by the {\it
drag along} of the metric under an infinitesimal active
diffeomorphism ${}^4g \mapsto {}^4\tilde g$ so that

\beq
 \bar \delta\, {}^4g_{\mu\nu}  \equiv {}^4\tilde g_{\mu\nu}(x) -
{}^4g_{\mu\nu}(x) =  -\Big( \xi_{\mu ;\nu}(x, {}^4g) + \xi_{\nu
;\mu}(x, {}^4g) \Big),
 \label{II7}
 \eeq

\noindent the solution $\xi_{\mu}(x, {}^4g)$ of these Killing-type
equations identifies a corresponding {\it passive} Bergmann-Komar
dynamical symmetry belonging to $Q$. We see that the new system of
coordinates $[x^{{'}\mu}]$ is identical to the {\it drag along} of
the old coordinate system, so that here we have made {\it
explicit} the merely implicit {\it dual view} quoted above.

\bigskip

This result should imply that all the {\it active diffeomorphisms
connected to the identity in ${}_ADiff\, M^4$} can be
reinterpreted as elements of a {\it non-normal} sub-group of {\it
generalized passive transformations in $Q$}. Clearly this
sub-group is disjoint from the sub-group ${}_PDiff\, M^4$: again,
this is possible because diffeomorphism groups do not possess a
canonical identity. However, let us recall that, unfortunately,
there is no viable mathematical treatment of the diffeomorphism
group in the large.

In conclusion, what is known as {\it 4-geometry}, or as {\it
Einstein (or on-shell, or dynamical) gravitational field}, is also
an equivalence class of solutions of Einstein's equations {\it
modulo} the dynamical symmetry transformations of ${}_ADiff\,
M^4$. Therefore, usually one finds the following statement
\cite{8a}

\beq
 {}^4Geom = {}^4Riem / {}_PDiff\, M^4 = {}^4Riem / Q' = {}^4Riem /
 {}_ADiff\, M^4.
 \label{II8}
 \eeq

It should be stressed, however, that the last two equalities hold
in the previously explained weak sense.

\medskip

It is clear that a parametrization of the 4-geometries should be
grounded on the two independent dynamical degrees of freedom of
the gravitational field. Within the framework of the Lagrangian
dynamics, however, no algorithm is known for evaluating the
observables of the gravitational field, viz. its two independent
degrees of freedom. The only result we know of is given in
Ref.\cite{9a} where, after a study of the index of Einstein's
equations, it is stated that the two degrees of freedom are
locally associated to {\it symmetric trace-free 2-tensors on
two-planes}, suggesting a connection with the Newman-Penrose
formalism \cite{15a}.

On the other hand, as we shall see in the next Section, it is the
Hamiltonian framework which has the proper tools to attack these
problems. Essentially, this is due to the fact that the
Hamiltonian methods allow to work {\it off-shell}, i.e., without
immediate transition to the space of solutions of Einstein's
equations. Thus the soldering to the above results is reached {\it
only} at the end of the canonical reduction, when the {\it
on-shell} restriction is made \footnote{Note nevertheless that
even at the Lagrangian level one can define {\it off-shell (or
kinematical) gravitational fields} defined as ${}^4Riem^{'} /
{}_PDiff\, M^4$, where ${}^4Riem^{'}$ are all the possible metric
tensors on $M^4$. Of course only the subset of solutions of
Einstein equations are Einstein gravitational fields.}.

\bigskip

Let us now make some remarks about the choice of coordinate
systems. On the one hand, it is clear from Eq.(\ref{II8}) that,
given a solution of Einstein's equations in a coordinate system,
its form in any other system, either ordinary or extended, can be
obtained by means of Eqs.(\ref{II3}) or (\ref{II4}). On the other
hand, in practice one looks for the most convenient coordinate
system for dealing with specific problems. This is always done by
imposing some conditions to be satisfied by the metric tensor in
the wanted coordinate system, so that such coordinate conditions
amount to a complete or partial breaking of general covariance. In
the variational approach A) these conditions are named {\it
Lagrangian gauge fixings} \footnote{As we shall see, in the
canonical formulation of general relativity one speaks of
Hamiltonian gauge fixings, which correspond to a fixation of the
coordinates of $M^4$ only {\it on-shell}. In particular, the
fixation of the 3-coordinates on a Cauchy surface are made by
imposing 3 gauge fixing constraints on the 3-metric. }. If we
start with Einstein's equations in an arbitrary coordinate system
$x^{\mu}$ of the atlas of $M^4$, the transition to the special
coordinate system $x^{{'} \mu}$, identified by a set of conditions
on the metric, may either correspond to an ordinary coordinate
transformation ({\it passive diffeomorphism}) $x^{{'} \mu} =
f^{\mu}(x)$ between two charts of the atlas of $M^4$ or, most
likely, to an extended transformation of the type (\ref{II4})
(passive re-interpretation of an {\it active diffeomorphism}).

\medskip

i) The usual search for exact solutions of Einstein's equations
relies on a choice of coordinates dictated by the assumed Killing
symmetries of the metric tensor, which are special metric
conditions.

\medskip

ii) The Lagrangian gauge fixing procedure amounts to the
determination of the inverse coordinate transformation $x^{\mu} =
h^{\mu}(x^{'})$ as a solution of Eq.(\ref{II3}) interpreted as a
partial differential equation for $h^{\mu}(x^{'})$, with the
metric ${}^4g^{'}_{\mu\nu}(x^{'})$ satisfying the required
conditions. Since the group of passive diffeomorphisms as well as
its extension (\ref{II4}) depend on four arbitrary functions, a
choice of either a specific coordinate system or a family of
coordinate systems has to be done by imposing $N$ suitable
functional conditions on the metric tensor (either $N = 4$ or $N
\leq 4$). Typical instantiations of this fact are the following:

\medskip

a) Algebraic Lagrangian gauge fixings:

a1) Family of {\it synchronous coordinates}:
${}^4g^{'}_{oi}(x^{'}) = 0$, $i = 1,2,3$; since $N = 3$, there is
a residual gauge freedom, namely the solution $h^{\mu}$ depends
upon an arbitrary function.

a2) Family of {\it 3-orthogonal coordinates}:
${}^4g^{'}_{ij}(x^{'}) = 0$, $i \not= j$; again there is a
residual gauge freedom depending upon an arbitrary function.
\medskip

b) Non-algebraic Lagrangian gauge fixings, in which the metric
${}^4g^{'}_{\mu\nu}(x^{'})$  is only restricted to be a solution
of partial differential equations, so that there is an extra
dependence upon new arbitrary functions:

b1) Family of {\it harmonic coordinates}: they are associated to
all the functional forms of ${}^4g^{'}_{\mu\nu}(x^{'})$ which
satisfy the four partial differential equations:
$\Gamma^{\alpha}_{\mu\nu}[{}^4g^{'}(x^{'})]\, {}^4g^{{'}\,
\mu\nu}(x^{'}) = 0$.

b2) Family of {\it Riemann normal coordinates} around a point
\cite{19a}: they are defined by asking that the geodesics
emanating from the point are straight lines.

\bigskip

Let us end this Section with a remark on general covariance that,
with the exception of Kretschmann \cite{20a}, is usually
considered a genuine and fundamental feature of general relativity
which can be extended to special relativity and Newton mechanics
only in a formal and artificial way\footnote{In Ref.\cite{13a} it
is shown that within {\it parametrized Minkowski theories} it is
possible to re-formulate the dynamics of isolated systems in
special relativity on arbitrary space-like hyper-surfaces that are
leaves of the foliation associated with an arbitrary 3+1 splitting
and also define a surface-forming congruence of accelerated
time-like observers. In these theories the embeddings
$z^{\mu}(\tau ,\sigma )$ of the space-like hyper-surfaces are new
configuration variables at the Lagrangian level. However, they are
{\it gauge variables} because the Lagrangian is invariant under
separate $\tau$- and $\vec \sigma$-reparametrizations (which are
diffeomorphisms). This form of {\it special relativistic general
covariance} implies the existence of four first class constraints
analogous to the super-hamitonian and super-momentum constraints
of ADM canonical gravity, which assure the independence of the
description from the choice of the 3+1 splitting.}.

\medskip

Let us remark that, in special relativity, the embedding $x^{\mu}
= z^{\mu}(\tau ,\vec \sigma )$ is usually described with respect
to the axes of an {\it instantaneous inertial observer} (see
Appendix A for the terminology concerning time-like observers)
chosen as origin of a {\it global inertial reference frame},
namely a congruence of time-like straight-lines parallel to the
time axis of the instantaneous inertial observer. More generally,
we can introduce (already in Minkowski space-time) a {\it global
non-inertial reference frame} defined as a congruence of time-like
world-lines, determined by a unit vector field, one of which is
selected as an {\it instantaneous non-inertial observer}
$X^{\mu}(\tau )$. This latter is then used as the centroid, origin
of the curvilinear 3-coordinate system $\sigma^r$, $r=1,2,3$, on
the simultaneity $\tau = const.$ 3-surfaces $\Sigma_{\tau}$, so
that the embeddings can be parametrized as $z^{\mu}(\tau ,\vec
\sigma ) = X^{\mu}(\tau ) + F^{\mu}(\tau ,\vec \sigma )$,
$F^{\mu}(\tau ,\vec 0) = 0$. See Ref.\cite{12a} for the definition
of the admissible embeddings in special relativity.

\medskip

Obviously, in curved space-times, globally inertial reference
frames do not exist (only local ones do, freely falling along
4-geodesics), but still we can safely use the notion of {\it
global non-inertial laboratory} provided that the topology of
$M^4$ is trivial. To every such frame a special global coordinate
chart $x^{\mu}$ in the atlas of $M^4$ can be associated.

\vfill\eject

\section{The ADM Hamiltonian Viewpoint and The Related Canonical Local
Symmetries.}

This Section provides the analysis of the Cauchy problem and the
counting of degrees of freedom within the framework of the ADM
canonical formulation of metric gravity \cite{21a}. Since we are
interested in a model of general relativity able to incorporate
the standard model of elementary particles and its extensions, and
since these models are a chapter of the theory of representations
of the Poincare' group on Minkowski space-time, we will consider
only non-compact, topologically trivial space-times. Moreover they
must be globally hyperbolic pseudo-Riemannian 4-manifolds $M^4$
asymptotically flat at spatial infinity, because only in this case
a Hamiltonian formulation is possible. Actually, unlike the
Lagrangian formulation, the Hamiltonian formalism requires a 3+1
splitting of $M^4$ and a global {\it mathematical time} function
$\tau$. This entails in turn a foliation of $M^4$ by space-like
hyper-surfaces $\Sigma_{\tau}$ \footnote{The 3-surfaces
$\Sigma_{\tau}$ are instances of {\it equal time surfaces
corresponding to a convention of synchronization of distant
clocks, a definition of 3-space and a determination of the one-way
velocity of light}, generalizing the customary Einstein convention
valid only in the inertial systems of special relativity. For a
discussion of this topic see Ref.\cite{12a}. The use of the
parameter $\tau$, labeling the leaves of the foliation, as an
evolution parameter corresponds to the {\it hyper-surface point of
view} of Ref.\cite{22a}. The {\it threading point of view} is
instead a description involving only a rotating congruence of
observers: since the latter is rotating, it is not surface-forming
(non-zero vorticity) and in each point we can only divide the
tangent space in the direction parallel to the 4-velocity and the
orthogonal complement (the local rest frame). On the other hand,
the {\it slicing point of view}, originally adopted in ADM
canonical gravity, uses two congruences: the non-rotating one with
the normals to $\Sigma_{\tau}$ as 4-velocity fields and a second
(rotating, non-surface-forming) congruence of observers, whose
4-velocity field is the field of time-like unit vectors determined
by the $\tau$ derivative of the embeddings identifying the leaves
$\Sigma_{\tau}$ (their so-called evolution vector field).
Furthermore, as Hamiltonian evolution parameter it uses the affine
parameter describing the world-lines of this second family of
observers.}({\it simultaneity Cauchy surfaces}, assumed
diffeomorphic to $R3$ so that any two points on them are joined by
a unique 3-geodesic), to be coordinatized by {\it adapted}
3-coordinates $\vec \sigma$ \footnote{An improper vector notation
is used throughout for the sake of simplicity.}.

\medskip

If $\tau$ is the mathematical time labeling these 3-surfaces,
$\Sigma_{\tau}$, and $\vec \sigma$ are 3-coordinates (with respect
to an arbitrary observer, a centroid $X^{\mu}(\tau )$, chosen as
origin) on them, then $\sigma^A = (\tau ,\vec \sigma )$ can be
interpreted as {\it Lorentz-scalar radar 4-coordinates} and the
surfaces $\Sigma_{\tau}$ are described by embedding functions
$x^{\mu} = z^{\mu}(\tau ,\vec \sigma ) = X^{\mu}(\tau ) +
F^{\mu}(\tau ,\vec \sigma )$, $F^{\mu}(\tau ,\vec 0) = 0$. In
these coordinates the metric is ${}^4g_{AB}(\tau ,\vec \sigma ) =
z^{\mu}_A(\tau ,\vec \sigma )\, {}^4g_{\mu\nu}(z(\tau ,\vec \sigma
))\, z^{\nu}_B(\tau ,\vec \sigma )$ [$z^{\mu}_A =
\partial z^{\mu} / \partial \sigma^A$]. Since the 3-surfaces
$\Sigma_{\tau}$ are {\it equal time} 3-spaces with all clocks
synchronized, the spatial distance between two equal-time events
will be $dl_{12} = \int_{12}\, dl\, \sqrt{{}^3g_{rs}(\tau ,\vec
\sigma (l))\, {{d\sigma^r(l)}\over {dl}}\, {{d\sigma^s(l)}\over
{dl}}}\,\,$ [$\vec \sigma (l)$ is a parametrization of the
3-geodesic $\gamma_{12}$ joining the two events on
$\Sigma_{\tau}$]. Moreover, by using test rays of light we can
define the {\it one-way} velocity of light between events on
different $\Sigma_{\tau}$'s. Therefore, the Hamiltonian
description has naturally built in the tools (essentially the 3+1
splitting) to make contact with experiments in a relativistic
framework, where simultaneity is a frame-dependent property. Let
us note that the manifestly covariant description using Einstein's
equations is the natural one for the search of exact solutions,
but is inadequate to describe experiments.

\bigskip

As shown in Ref.\cite{13a}, with this formulation the so-called
{\it problem of time} can be treated in such a way that in
presence of matter and in the special-relativistic limit of
vanishing Newton constant, one recovers the parametrized Minkowski
theories, quoted at the end of the previous Section, equipped with
a {\it global time} $\tau$. A canonical formulation with
well-defined Poisson brackets requires in addition the
specification of suitable {\it boundary conditions} at spatial
infinity and a definite choice of the functional space for the
fields. While the problem of the boundary conditions constitutes
an intriguing issue within the Lagrangian approach, the
Hamiltonian one is more easy to treat. Even if we shall consider
only metric gravity, let us remark that with the inclusion of
fermions it is natural to resolve the metric tensor in terms of
cotetrad fields \cite{23a} [${}^4g_{\mu\nu}(x) = E^{(\alpha
)}_{\mu}(x)\, \eta_{(\alpha )(\beta )}\, E^{(\beta )}_{\nu}(x)$;
$\eta_{(\alpha )(\beta )}$ is the flat Minkowski metric in
Cartesian coordinates] and to reinterpret the gravitational field
as a {\it theory of time-like observers endowed with tetrads},
whose dynamics is controlled by the ADM action thought as a
function of the cotetrad fields.

\medskip

Only the aspects important to our program will be reviewed here.
The reader is referred to Ref.\cite{13a} for the relevant
notations and the general technical development of the Hamiltonian
description of metric gravity, which requires the use of
Dirac-Bergmann \cite{24a,25a,26a,27a,28a} theory of constraints
(see Refs.\cite{29a,30a} for updated reviews). We use a Lorentzian
signature $\epsilon\, (+---)$, with $\epsilon = \pm 1$ according
to particle physics and general relativity conventions,
respectively.

\subsection{ADM Action, Asymptotic Symmetries and Boundary
Conditions.}

We start off with replacement of the ten components
${}^4g_{\mu\nu}$ of the 4-metric tensor by the configuration
variables of ADM canonical gravity: the {\it lapse} $N(\tau , \vec
\sigma )$ and {\it shift} $N_r(\tau ,\vec \sigma )$ functions and
the six components of the {\it 3-metric tensor} on
$\Sigma_{\tau}$, ${}^3g_{rs}(\tau , \vec \sigma )$. We have $
{}^4g_{AB}=\left( \begin{array}{ll} {}^4g_{\tau\tau}= \epsilon
(N^2-{}^3g_{rs}N^rN^s)& {}^4g_{\tau s}=- \epsilon \, {}^3g_{su}N^u
\\ {}^4g_{\tau r}=- \epsilon \, {}^3g_{rv}N^v&
{}^4g_{rs}=-\epsilon \, {}^3g_{rs} \end{array} \right)$.
Einstein's equations are then recovered as the Euler-Lagrange
equations of the ADM action

\bea
  S_{ADM}&=&\int d\tau \, L_{ADM}(\tau )= \int d\tau d^3\sigma
{\cal L}_{ADM}(\tau ,\vec \sigma )=\nonumber \\
 &=&- \epsilon k\int_{\triangle \tau}d\tau  \, \int d^3\sigma \, \lbrace
\sqrt{\gamma} N\, [{}^3R+{}^3K_{rs}\, {}^3K^{rs}-({}^3K)^2]\rbrace
(\tau ,\vec \sigma ),
 \label{III1}
 \eea

\noindent which differs from Einstein-Hilbert action (\ref{II1})
by a suitable surface term. Here ${}^3K_{rs}$ is the extrinsic
curvature of $\Sigma_{\tau}$, ${}^3K$ its trace, and ${}^3R$ the
3-curvature scalar.

\bigskip

Besides the ten configuration variables listed above, the ADM
functional phase space $\Gamma_{20}$ is {\it coordinatized} by ten
canonical momenta ${\tilde \pi}^N(\tau ,\vec \sigma )$, ${\tilde
\pi}^r_{\vec N}(\tau ,\vec \sigma )$, ${}^3{\tilde \Pi}^{rs}(\tau
, \vec \sigma )$. Such canonical variables, however, are not
independent since they are restricted to the {\it constraint
sub-manifold} $\Gamma_{12}$ by the eight {\it first class}
constraints [${}^3G_{rstw} = {}^3g_{rt}\, {}^3g_{sw} +
{}^3g_{rw}\, {}^3g_{st} - {}^3g_{rs}\, {}^3g_{tw}$ is the
Wheeler-DeWitt super-metric]

\bea
 {\tilde \pi}^N(\tau ,\vec \sigma ) &\approx& 0 ,\nonumber \\
 {\tilde \pi}^r_{\vec N}(\tau ,\vec \sigma ) &\approx& 0,\nonumber
 \\
 &&{}\nonumber \\
 {\tilde {\cal H}}(\tau ,\vec \sigma )&=&\epsilon [k\sqrt{\gamma}\,
{}^3R-{1\over {2k \sqrt{\gamma}}} {}^3G_{rsuv}\, {}^3{\tilde
\Pi}^{rs}\, {}^3{\tilde \Pi}^{uv}] (\tau ,\vec \sigma ) \approx
0,\nonumber \\
 {}^3{\tilde {\cal H}}^r(\tau ,\vec \sigma )&=&-2\, {}^3{\tilde
\Pi}^{rs}{}_{| s} (\tau ,\vec \sigma )=-2[\partial_s\, {}^3{\tilde
\Pi}^{rs}+{}^3\Gamma^r_{su} {}^3{\tilde \Pi}^{su}](\tau ,\vec
\sigma ) \approx 0.
 \label{III2}
  \eea

While the first four are {\it primary} constraints, the remaining
four are the super-hamiltonian and super-momentum  {\it secondary}
constraints arising from the requirement that the primary
constraints be constant in $\tau$. More precisely, this
requirement guarantees that, once we have chosen the initial data
inside the constraint sub-manifold $\Gamma_{12}(\tau_o)$
corresponding to a given initial Cauchy surface $\Sigma_{\tau_o}$,
the time evolution does not take them out of the constraint
sub-manifolds $\Gamma_{12}(\tau)$, for $\tau > \tau_o$.

The evolution in $\tau$ is ruled by the Hamilton-Dirac Hamiltonian

\beq
 H_{(D)ADM}=\int d^3\sigma \, \Big[ N\, {\tilde {\cal H}}+N_r\,
{}^3{\tilde {\cal H}}^r + \lambda_N\, {\tilde \pi}^N + \lambda
^{\vec N}_r\, {\tilde \pi}^r_{\vec N}\Big](\tau ,\vec \sigma )
\approx 0,
 \label{III3}
 \eeq

\noindent where $\lambda_N(\tau , \vec \sigma )$ and
$\lambda^r_{\vec N}(\tau ,\vec \sigma )$ are {\it arbitrary Dirac
multipliers} in front of the primary constraints\footnote{These
are four {\it velocity functions} (gradients of the metric tensor)
which are not determined by Einstein's equations. }. The resulting
hyperbolic system of Hamilton-Dirac equations has the same
solutions of the non-hyperbolic system of (Lagrangian) Einstein's
equations with the same boundary conditions. Let us stress that
Hamiltonian hyperbolicity is explicitly paid by the arbitrariness
of the Dirac multipliers\footnote{Of course this is just the
Hamiltonian counterpart of the "{\it indeterminateness}" or the
so-called "{\it indeterminism}" surfacing in what Einstein called
Hole Argument ("Lochbetrachtung") in 1915-1916 \cite{1a}.}.

\bigskip

At this point let us see the further conditions to be required
with respect to the above standard ADM formulation.

Additional requirements \cite{13a} on the Cauchy and simultaneity
3-surfaces $\Sigma_{\tau}$ induced by particle physics are:

i) Each $\Sigma_{\tau}$ must be a Lichnerowitz 3-manifold
\cite{31a}, namely it must admit an involution so that a
generalized Fourier transform can be defined and the notion of
positive and negative frequencies can be introduced (otherwise the
notion of particle cannot be properly defined, like it happens in
quantum field theory in arbitrary curved space-times \cite{32a}).

ii) Both the metric tensor and the fields of the standard model of
elementary particles must belong to the same family of suitable
weighted Sobolev spaces so that there are no Killing vector fields
on space-time (this avoids the cone-over-cone structure of
singularities in the space of metrics) and no Gribov ambiguity
(either gauge symmetries or gauge copies \cite{33a}) in the
particle sectors; in both cases no well defined Hamiltonian
description could be available.

iii) Space-time must be {\it asymptotically flat at spatial
infinity} and satisfying boundary conditions there in a way
independent of the direction (in analogy to what is required for
the defining non-Abelian charges in Yang-Mills theory \cite{33a}).
This eliminates the {\it supertranslations} (i.e., the obstruction
to define angular momentum in general relativity) and reduces the
{\it spi group} of asymptotic symmetries to the ADM Poincare'
group. The constant ADM Poincare' generators should become the
standard conserved Poincare' generators of the standard model of
elementary particles when gravity is turned off and space-time
(modulo a possible renormalization of the ADM energy to subtract
an infinite term coming from its dependence on both $G$ and $1/G$)
becomes Minkowskian\footnote{Incidentally, this is the first
example of consistent {\it deparametrization} of general
relativity. In presence of matter we get the description of matter
in Minkowski space-time foliated with the space-like hyper-planes
orthogonal to the total matter 4-momentum (Wigner hyper-planes
intrinsically defined by matter isolated system). Of course, in
closed space-times, the ADM Poincare' charges do not exist and the
special relativistic limit is lost.}. As a consequence, the {\it
admissible foliations} of the space-time must have the
simultaneity surfaces $\Sigma_{\tau}$ tending in a
direction-independent way to Minkowski space-like hyper-planes at
spatial infinity, where they must be orthogonal to the ADM
4-momentum. Now, these are exactly the conditions satisfied by the
Christodoulou-Klainermann space-times \cite{9a} , which are near
Minkowski space-time in a norm sense and have a {\it rest-frame}
condition of zero ADM 3-momentum. The hyper-surfaces
$\Sigma_{\tau}$ define the {\it rest frame} of the $\tau$-slice of
the universe and admit  {\it asymptotic inertial observers} to be
identified with the {\it fixed stars} (this also defines the {\it
standard of rotations} the spatial precession of gyroscopes is
referred to)\footnote{These properties are concretely enforced
\cite{13a} by using  a technique introduced by Dirac \cite{24a}
for the selection of space-times admitting asymptotically flat
4-coordinates at spatial infinity. Dirac's method brings to an
enlargement of the ADM phase space, subsequently reduced to the
standard one by adding suitable constraints, as shown explicitly
in Ref.\cite{13a}. As a consequence the admissible embeddings of
the simultaneity leaves $\Sigma_{\tau}$ have the following
direction-independent limit at spatial infinity: $z^{\mu}(\tau
,\vec \sigma ) = X^{\mu}(\tau ) + F^{\mu}(\tau ,\vec \sigma )
\rightarrow_{|\vec \sigma | \rightarrow \infty}\,
X^{\mu}_{(\infty)}(0) + \epsilon^{\mu}_A\, \sigma^A =
X^{\mu}_{(\infty)}(\tau ) + \epsilon^{\mu}_r\, \sigma^r$. Here
$X^{\mu}_{(\infty)}(\tau ) = X^{\mu}_{(\infty)}(0) +
\epsilon^{\mu}_{\tau}\, \tau$ is just the world-line of an
asymptotic inertial observer having $\tau$ as proper time and
$\epsilon^{\mu}_A$ denotes an asymptotic constant tetrad with
$\epsilon^{\mu}_{\tau}$ parallel to the ADM 4-momentum (it is
orthogonal to the asymptotic space-like hyper-planes). Such
inertial observers corresponding to the {\it fixed stars} can be
endowed with a spatial triad ${}^3e^r_{(a)} = \delta^r_{(a)}$,
$a=1,2,3$. Then the asymptotic spatial triad ${}^3e^r_{(a)}$ can
be transported in a dynamical way (on-shell) by using the
Sen-Witten connection \cite{34a} (it depends on the extrinsic
curvature of the $\Sigma_{\tau}$'s) in the Frauendiener
formulation \cite{35a} in {\it every point} of $\Sigma_{\tau}$,
where it becomes a well defined triad ${}^3e_{(a)}^{(WSW) r}(\tau
,\vec \sigma )$. This defines a {\it local compass of inertia}, to
be compared with the local gyroscopes (whether Fermi-Walker
transported or not). The Wigner-Sen-Witten (WSW) local compass of
inertia consists in pointing to the fixed stars with a telescope.
It is needed in a satellite like Gravity Probe B to detect the
frame-dragging (or gravito-magnetic Lense-Thirring effect) of the
inertial frames by means of the rotation of a FW transported
gyroscope.\hfill\break Finally from Eq.(12.8) of Ref.\cite{13a} we
get the following set of partial differential equations for the
determination of the embedding $x^{\mu} = z^{\mu}(\tau ,\vec
\sigma )$ ($x^{\mu}$ is an arbitrary 4-coordinate system in which
the asymptotic hyper-planes of the $\Sigma_{\tau}$'s have
$\epsilon^{\mu}_A$ as asymptotic tetrad)\break \hfill\break
\begin{eqnarray*}
 z^{\mu}(\tau ,\vec \sigma ) &=& X^{\mu}_{(\infty)}(0) + F^A(\tau
 ,\vec \sigma )\, {{\partial z^{\mu}(\tau ,\vec \sigma )}\over
 {\partial \sigma^A}},\nonumber \\
 &&{}\nonumber \\
 F^{\tau}(\tau ,\vec \sigma ) &=& {{-\epsilon\, \tau}\over
 {-\epsilon + n(\tau ,\vec \sigma )}},\nonumber \\
 F^r(\tau ,\vec \sigma ) &=& \sigma^r + [{}^3e_{(a)}^{(WSW) r}(\tau
 ,\vec \sigma )\, - \delta^r_{(a)} ]\, \delta_{(a)s}\, \sigma^s + {{\epsilon\, n^r(\tau
 ,\vec \sigma )}\over {-\epsilon + n(\tau ,\vec \sigma )}}.
 \end{eqnarray*} }. Another interesting point is that this class of space-times
admits an {\it asymptotic Minkowski metric} (asymptotic
background) which allows to define weak gravitational field
configurations and background-independent gravitational waves
\cite{36a} that do not require {\it splitting of the metric} in a
background term plus a perturbation (and without being a bimetric
theory of gravity).
\medskip

As shown in Ref.\cite{13a}, a consistent treatment of the boundary
conditions at spatial infinity requires the explicit separation of
the {\it asymptotic} part of the lapse and shift functions from
their {\it bulk} part: $N(\tau ,\vec \sigma ) = N_{(as)}(\tau
,\vec \sigma ) + n(\tau , \vec \sigma )$, $N_r(\tau ,\vec \sigma )
= N_{(as)r}(\tau ,\vec \sigma ) + n_r(\tau , \vec \sigma )$, with
$n$ and $n_r$ tending to zero at spatial infinity in a
direction-independent way\footnote{We would like to recall that
Bergmann \cite{14a} made the following critique of general
covariance: it would be desirable to restrict the group of
coordinate transformations (space-time diffeomorphisms) in such a
way that it could contain an invariant sub-group describing the
coordinate transformations that change the frame of reference of
an outside observer; the remaining coordinate transformations
would be like the gauge transformations of electromagnetism. This
is just what is done here by the redefinition of the lapse and
shift functions after separating out their asymptotic part. In
this way, {\it preferred} inertial asymptotic coordinate systems
are selected that can be identified as fixed stars.} . On the
contrary, $N_{(as)}(\tau ,\vec \sigma ) = - \lambda_{\tau}(\tau )
- {1\over 2}\, \lambda_{\tau u}(\tau )\, \sigma^u$ and
$N_{(as)r}(\tau ,\vec \sigma ) = - \lambda_{r}(\tau ) - {1\over
2}\, \lambda_{r u}(\tau )\, \sigma^u$. In the
Christodoulou-Klainermann space-times \cite{9a} we have
$N_{(as)}(\tau ,\vec \sigma ) = \epsilon$, $N_{(as) r}(\tau ,\vec
\sigma ) = 0$.

\bigskip

Recall that the evolution is parametrized  by the mathematical
parameter $\tau$ of the adapted coordinate system $(\tau ,\vec
\sigma )$ on $M^4$, which labels the surfaces $\Sigma_{\tau}$. As
shown in Ref.\cite{13a}, the Hamiltonian ruling the evolution is
the {\it weak ADM energy} \cite{37a} (the volume form $E_{ADM}$).
As shown by DeWitt \cite{38a}, this is a consequence of the fact
that {\it in non-compact space-times the weakly vanishing ADM
Dirac Hamiltonian (\ref{III3}) has to be modified with a suitable
surface term in order to have functional derivatives, Poisson
brackets and Hamilton equations mathematically well defined}.

It follows, therefore, that the boundary conditions of this model
of general relativity imply that the real Dirac Hamiltonian
is\footnote{As shown in Ref.\cite{13a}, the correct treatment of
the boundary conditions leads to rewrite Eqs.(\ref{III3}) and
(\ref{III4}) in terms of $n$ and $n_r$. Moreover the momenta
${\tilde \pi}^N$,  ${\tilde \pi}^r_{\vec N}$ should be always
replaced by  ${\tilde \pi}^n$, ${\tilde \pi}^r_{\vec n}$.}

\beq
 H_D = E_{ADM} + H_{(D)ADM} \approx E_{ADM},
 \label{III4}
 \eeq

\noindent and this entails that {\it an effective evolution takes
place in mathematical time $\tau$}\footnote{As we shall see, the
super-hamiltonian constraint is only the generator of the gauge
transformations connecting different admissible 3+1 splittings of
space-time and {\it has nothing to do with the temporal evolution}
(no Wheeler-DeWitt interpretation).}, and that a non-vanishing
Hamiltonian survives in the reduced phase space of the intrinsic
degrees of freedom (no frozen reduced phase space picture).

The weak ADM energy, and also the other nine asymptotic weak
Poincare' charges ${\vec P}_{ADM}$, $J^{AB}_{ADM}$, {\it are
Noether constants of the motion whose numerical value has to be
given as part of the boundary conditions}. The numerical value of
$E_{ADM}$ is the mass of the $\tau$-slice of the universe, while
$J^{rs}_{ADM}$ gives the value of the spin of the universe. Since,
in our case, space-time is of the Christodoulou-Klainermann type
\cite{9a}, the ADM 3-momentum has to vanish. This implies three
first class constraints

\beq
 {\vec P}_{ADM} \approx 0,
 \label{III5}
 \eeq

\noindent which identify the {\it rest frame of the universe}. As
shown in Ref.\cite{13a}, the natural gauge fixing to these three
constraints is the requirement the the ADM boosts vanish: $J^{\tau
r}_{ADM} \approx 0$. In this way we decouple from the universe its
3-center of mass \footnote{This is equivalent to a choice of the
centroid $X^{\mu}(\tau )$ [or of the asymptotic one
$X^{\mu}_{(\infty)}(\tau )$], origin of the 3-coordinates on each
$\Sigma_{\tau}$.} and only {\it relative motions} survive,
recovering a Machian flavour.

\subsection{Hamiltonian Gauge Transformations.}

At this point a number of important questions must be clarified.
When used as generators of canonical transformations, the eight
first class constraints will map points of the constraint surface
to points on the same surface. We shall say that they generate the
infinitesimal transformations of the {\it off-shell Hamiltonian
gauge group} ${\cal G}_8$ \footnote{Note that the off-shell
Hamiltonian gauge transformations are {\it local Noether
transformations} (second Noether theorem) under which the ADM
Lagrangian (\ref{III1}) is {\it quasi-invariant}.}. The action of
${\cal G}_8$ gives rise to a {\it Hamiltonian gauge orbit} through
each point of the constraint sub-manifold $\Gamma_{12}$. Every
such orbit is parametrized by eight phase space functions - namely
the independent {\it off-shell Hamiltonian gauge variables} -
conjugated to the first class constraints. We are left thereby
with a pair of conjugate canonical variables, the {\it off-shell
DO}, which are the only {\it Hamiltonian gauge-invariant and
deterministically ruled} quantities. The same counting of degrees
of freedom of the Lagrangian approach is thus obtained. Finally,
let us stress here, in view of the later discussion, that both the
off-shell Christoffel symbols and the off-shell Riemann tensor can
be read as functions of both the off-shell DO and the Hamiltonian
gauge variables.

\bigskip

The eight infinitesimal off-shell Hamiltonian gauge
transformations have the following interpretation\cite{13a}:

i) those generated by the four primary constraints modify the
lapse and shift functions: these in turn determine how densely the
space-like hyper-surfaces $\Sigma_{\tau}$ are distributed in
space-time and also the conventions to be {\it pre-fixed} on each
$\Sigma_{\tau}$ about gravito-magnetism (see Section IV of
Ref.\cite{36a} for its dependence upon the choice of gauge, i.e.
on-shell of the 4-coordinates);

ii) those generated by the three super-momentum constraints induce
a transition on $\Sigma_{\tau}$ from a given 3-coordinate system
to another one;

iii) that generated by the super-hamiltonian constraint induces a
transition from a given 3+1 splitting of $M^4$ to another, by
operating normal deformations \cite{39a} of the space-like
hyper-surfaces\footnote{Note that in {\it compact} space-times the
super-hamiltonian constraint is usually interpreted as generator
of the evolution in some {\it internal time}, either like York's
internal {\it extrinsic} time or like Misner's internal {\it
intrinsic} time. In this paper instead the super-hamiltonian
constraint is the generator of those Hamiltonian gauge
transformations which imply that the description is independent of
the choice of the allowed 3+1 splitting of space-time: {\it this
is the proper answer to the criticisms raised against the phase
space approach on the basis of its lack of manifest covariance}.
}.

iv) those generated by the three rest-frame constraints
(\ref{III5}) can be interpreted as a change of centroid to be used
as origin of the 3-coordinates.

\medskip

As a consequence, the whole set of Hamiltonian off-shell gauge
transformations contains also a change of the {\it global
non-inertial space-time laboratory} and its associated
coordinates.

\bigskip

Making the quotient of the constraint hyper-surface with respect
to the off-shell Hamiltonian gauge transformations by defining
$\Gamma_4 = \Gamma_{12} / {\cal G}_8$, we obtain the so-called
{\it reduced off-shell conformal super-space}. Each of its points,
i.e. a {\it Hamiltonian off-shell (or kinematical) gravitational
field}, is an off-shell equivalence class, called an {\it
off-shell conformal 3-geometry}, for the space-like hyper-surfaces
$\Sigma_{\tau}$: note that, since it contains all the off-shell
4-geometries connected by Hamiltonian gauge transformations, {\it
it is not a 4-geometry}.

\bigskip

An important digression is in order here. The space of parameters
of the off-shell gauge group ${\cal G}_8$ contains eight arbitrary
functions. Four of them are the Dirac multipliers $\lambda_N(\tau
,\vec \sigma )$, $\lambda_r^{\vec N}(\tau ,\vec \sigma )$ of
Eqs.(\ref{III3}), while the other four are functions $ \alpha
(\tau ,\vec \sigma )$, $\alpha_r(\tau ,\vec \sigma )$ which
generalize the lapse and shift functions in front of the secondary
constraints in Eqs.(\ref{III3}) \footnote{In Ref. \cite{8a} they
are called {\it descriptors} and written in the form $\alpha = N\,
\xi$, $\alpha^r = {}^3g^{rs}\, \alpha_s = \xi^r \pm N^r\, \xi$. }.
These arbitrary functions correspond to the eight local Noether
symmetries under which the ADM action is quasi-invariant.

On the other hand, from the analysis of the dynamical symmetries
of the Hamilton equations (equivalent to Einstein's equations), it
turns out (see Refs.\cite{40a,41a}) that {\it on-shell} only a
sub-group ${\cal G}_{4\, dyn}$ of ${\cal G}_8$ survives, depending
on  four arbitrary functions. But in the present context, a
crucial result for our subsequent discussion is that a further
subset, denoted by ${\cal G}_{4\, P} \subset {\cal G}_{4\, dyn}$,
can be identified within the sub-group ${\cal G}_{4\, dyn}$:
precisely the subset corresponding to the phase space counterparts
of those passive diffeomorphisms which are {\it projectable} to
phase space. On the other hand, as already said, Einstein's
equations have $Q$ as the largest group of dynamical symmetries
and, even if irrelevant to the local Noether symmetries of the ADM
action, the existence of this larger group is a fundamental
mathematical premise to our second paper II. In order to take it
into account in the present context, the parameter space of ${\cal
G}_8$ must be enlarged to arbitrary functions depending also on
the 3-metric, $\lambda_N(\tau ,\vec \sigma ) \mapsto
\lambda_N(\tau ,\vec \sigma , {}^3g_{rs}(\tau ,\vec \sigma ))$,
... , $\alpha_r(\tau ,\vec \sigma ) \mapsto \alpha_r(\tau ,\vec
\sigma , {}^3g_{rs}(\tau ,\vec \sigma ))$. Then, the restriction
of this enlarged gauge group to the {\it dynamical symmetries} of
Hamilton equations defines an extended group ${\tilde {\cal
G}}_{4\, dyn}$ which, under inverse Legendre transformation,
defines a new {\it non-normal} sub-group $Q_{can}$ of the group
$Q$ (see Ref.\cite{8a}). But now, the remarkable and fundamental
point is that $Q_{can}$ {\it contains both active and passive
diffeomorphisms}. In particular:
\bigskip

i) the intersection $Q_{can} \cap {}_PDiff\, M^4$ identifies the
space-time passive diffeomorphisms which, respecting the 3+1
splitting of space-time, are {\it projectable} to ${\cal G}_{4\,
P}$ in phase space;

ii) the remaining elements of $Q_{can}$ are the {\it projectable}
subset of active diffeomorphisms in their passive view.\bigskip

\noindent This entails that, as said in Ref.\cite{8a},
Eq.(\ref{II8}) may be completed with

\beq
 {}^4Geom = {}^4Riem / Q_{can}.
  \label{III6}
   \eeq

\bigskip

In conclusion, the real gauge group acting on the space of the
solutions of the Hamilton-Dirac equations  is the {\it on-shell
extended Hamiltonian gauge group} ${\tilde {\cal G}}_{4\, dyn}$
and the on-shell equivalence classes obtained by making the
quotient with respect to it eventually coincide with the on-shell
4-geometries of the Lagrangian theory. Therefore, the {\it
Hamiltonian Einstein (or on-shell, or dynamical) gravitational
fields} coincide with the Lagrangian Einstein (or {\it on-shell},
or {\it dynamical}) gravitational fields.\medskip

Let us remark that, while it is known how to formulate an initial
value problem for the partial differential equations of the
Hamiltonian theory in a complete Hamiltonian gauge and how to
connect the problems in different gauges by using on-shell
transformations in $Q_{can}$, no mathematical technique is known
for dealing with active diffeomorphisms in $Q'$ but not in
$Q_{can}$ in connection to the Cauchy problem within the framework
of abstract differential geometry. As already said, for the
configurational Einstein equations a technique, mimicking the
Hamiltonian treatment, does exist and the Cauchy problems in
different 4-coordinate systems are connected by transformations in
${}_PDiff\, M^4$.

\bigskip

This is the way in which {\it passive} space-time diffeomorphisms,
under which the Hilbert action is invariant, are reconciled {\it
on-shell} with the allowed Hamiltonian gauge transformations
adapted to the 3+1 splittings of the ADM formalism. Furthermore,
our analysis of the Hamiltonian gauge transformations and their
Legendre counterparts gives an extra {\it bonus}: namely that the
on-shell phase space extended gauge transformations include also
symmetries that are images of {\it active} space-time
diffeomorphsms. The basic relevance of this result for a deep
understanding of the so-called Hole Argument will appear fully in
paper II.

\bigskip

\subsection{The Shanmugadhasan Canonical Transformation and
the Canonical Reduction.}

Having clarified these important issues, let us come back to the
canonical reduction. The off-shell freedom corresponding to the
eight independent types of Hamiltonian gauge transformations is
reduced on-shell to four types like in the case of ${}_PDiff\,
M^4$: precisely the transformations in [$Q_{can} \cap {}_PDiff\,
M^4$]. At the off-shell level, this property is manifest by the
circumstance that the original Dirac Hamiltonian contains {\it
only} 4 arbitrary Dirac multipliers and that the {\it correct
gauge-fixing procedure} \cite{42a,13a} starts by giving {\it only}
the four gauge fixing to the secondary constraints. The gauge
fixing functions must satisfy the {\it orbit conditions} ensuring
that each gauge orbit is intersected only in one point by the
gauge fixing surface (locally this requires a non-vanishing
determinant of the Poisson brackets of the gauge functions with
the secondary constraints). Then, the requirement of time
constancy generates the four gauge fixing constraints to the
primary constraints, while time constancy of such secondary gauge
fixings leads to the determination of the four Dirac
multipliers\footnote{This agrees with the results of
Ref.\cite{43a} according to which the {\it projectable} space-time
diffeomorphisms depend only on four arbitrary functions and their
time derivatives.}. Since the original constraints plus the above
eight gauge fixing constraints form a second class set, it is
possible to introduce the associated {\it Dirac brackets} and
conclude the canonical reduction by realizing an off-shell reduced
phase space $\Gamma_4$. Of course, once we reach a {\it completely
fixed Hamiltonian gauge} (a copy of $\Gamma_4$), general
covariance is completely broken. Finally, recall that a completely
fixed Hamiltonian gauge is equivalent {\it on-shell} to a {\it
definite choice of the space-time 4-coordinates} on $M^4$, within
the Lagrangian viewpoint \cite{40a,41a}.

\bigskip

In order to visualize the meaning of the various types of degrees
of freedom\footnote{This visualization remains only implicit in
the conformal Lichnerowicz-York approach \cite{44a,45a,46a,47a}.}
we need the construction of a {\it Shanmugadhasan canonical basis}
\cite{10a} of metric gravity having the following structure ($\bar
a =1,2$ are non-tensorial indices of the DO\footnote{Let us recall
that the DO are in general neither tensors nor invariants under
space-time diffeomorphisms. Therefore their (unknown) functional
dependence on the original variables changes (off-shell) with the
gauge and, therefore, (on-shell) with the 4-coordinate system.}
$r_{\bar a}$, $\pi_{\bar a}$) with

 \bea
\begin{minipage}[t]{3cm}
\begin{tabular}{|l|l|l|} \hline
$n$ & $n_r$ & ${}^3g_{rs}$ \\ \hline ${\tilde \pi}^n \approx 0$ &
${\tilde \pi}_{\vec n}^r \approx 0$ & ${}^3{\tilde \Pi}^{rs}$ \\
\hline
\end{tabular}
\end{minipage} &&\hspace{2cm} {\longrightarrow \hspace{.2cm}} \
\begin{minipage}[t]{4 cm}
\begin{tabular}{|ll|l|l|l|} \hline
$n$ & $n_r$ & $\xi^{r}$ & $\phi$ & $r_{\bar a}$\\ \hline
 ${\tilde \pi}^n \approx 0$ & ${\tilde \pi}_{\vec n}^r \approx 0$
& ${\tilde \pi}^{{\vec {\cal H}}}_r \approx 0$ &
 $\pi_{\phi}$ & $\pi_{\bar a}$ \\ \hline
\end{tabular}
\end{minipage} \nonumber \\
 &&{}\nonumber \\
&& {\longrightarrow \hspace{.2cm}} \
\begin{minipage}[t]{4 cm}
\begin{tabular}{|ll|l|l|l|} \hline
$n$ & $n_r$ & $\xi^{r}$ & $Q_{\cal H} \approx 0$ & $r^{'}_{\bar
a}$\\ \hline
 ${\tilde \pi}^n \approx 0$ & ${\tilde \pi}^r_{\vec n} \approx 0$
& ${\tilde \pi}^{{\vec {\cal H}}}_r \approx 0$ &
 $\Pi_{\cal H}$ & $\pi^{'}_{\bar a}$ \\ \hline
\end{tabular}
\end{minipage}.
 \label{III7}
 \eea

\noindent It is seen that we need a sequence of two canonical
transformations.\bigskip

a) The first transformation replaces seven first-class constraints
with as many Abelian momenta ($\xi^r$ are the gauge parameters,
namely coordinates on the group manifold,  of the passive
3-diffeomorphisms generated by the super-momentum constraints) and
introduces the conformal factor $\phi$ of the 3-metric as the
configuration variable to be determined by the super-hamiltonian
constraint\footnote{Recall that the {\it strong} ADM energy is the
flux through the surface at spatial infinity of a function of the
3-metric only, and it is weakly equal to the {\it weak} ADM energy
(volume form) which contains all the dependence on the ADM
momenta. This implies \cite{13a} that the super-hamiltonian
constraint must be interpreted as the equation ({\it Lichnerowicz
equation}) that uniquely determines the {\it conformal factor}
$\phi = ( det\, {}^3g )^{1/12}$ of the 3-metric as a functional of
the other variables. This means that the associated gauge variable
is the {\it canonical momentum $\pi_{\phi}$ conjugate to the
conformal factor}: this latter carries information about the
extrinsic curvature of $\Sigma_{\tau}$. It is just this variable,
and {\it not} York's time, that parametrizes the {\it normal}
deformation of the embeddable space-like hyper-surfaces
$\Sigma_{\tau}$. As a matter of fact, a gauge fixing for the
super-hamiltonian constraint, i.e. a choice of a particular 3+1
splitting, is done by fixing the momentum $\pi_{\phi}$ conjugate
to the conformal factor. This shows the dominant role of the
conformal 3-geometries in the determination of the physical
degrees of freedom, just as in the Lichnerowicz-York conformal
approach.}. Note that the final gauge variable, namely the
momentum $\pi_{\phi}$ conjugate to the conformal factor, is the
only gauge variable of momentum type: it plays the role of a {\it
time} variable, so that the Lorentz signature of space-time is
made manifest by the Shanmugadhasan transformation in the set of
gauge variables $(\pi_{\phi}; \xi^r)$; this makes the difference
with respect to the proposals of Refs.\cite{48a,49a}. More
precisely, the first canonical transformation should be called a
{\it quasi-Shanmugadhasan } transformation, because nobody has
succeeded so far in Abelianizing the super-hamiltonian constraint.
Note furthermore that this transformation is a {\it point}
canonical transformation, whose inverse is known as a consequence
of the effect of finite gauge transformations (see Ref.\cite{23a}
for the case of tetrad gravity) .

b) The second canonical transformation would be instead a {\it
complete Shanmugadhasan} transformation, where $Q_{{\cal H}}(\tau
,\vec \sigma ) \approx 0$ would denote the Abelianization of the
super-hamiltonian constraint\footnote{If $\tilde \phi [r_{\bar a},
\pi_{\bar a}, \xi^r, \pi_{\phi}]$ is the solution of the
Lichnerowicz equation, then $Q_{{\cal H}}=\phi - \tilde \phi
\approx 0$. Other forms of this canonical transformation should
correspond to the extension of the York map \cite{50a} to
asymptotically flat space-times: in this case the momentum
conjugate to the conformal factor would be just York time and one
could add the maximal slicing condition as a gauge fixing. Again,
however, nobody has been able so far to build a York map
explicitly.}. The variables $n$, $n_r$, $\xi^r$, $\Pi_{\cal H}$
are the final {\it Abelianized Hamiltonian gauge variables}, while
$r^{'}_{\bar a}$, $\pi^{'}_{\bar a}$ are the final DO.
\bigskip

In absence of explicit solutions of the Lichnerowicz equation, the
best we can do is to construct the {\it quasi-Shanmugadhasan}
transformation. On the other hand, such transformation has the
remarkable property that, in the {\it special gauge}
$\pi_{\phi}(\tau ,\vec \sigma ) \approx 0$, the variables $r_{\bar
a}$, $\pi_{\bar a}$ form a canonical basis of off-shell DO for the
gravitational field {\it even if} the solution of the Lichnerowicz
equation is not known.

\bigskip

Let us stress the important fact that the Shanmugadhasan canonical
transformation is a {\it highly non-local}
transformation\footnote{This feature has a Machian flavor,
although in a non-Machian context: with or without matter, the
whole 3-space is involved in the definition of the observables.
Furthermore, these space-times allow the separation \cite{13a} of
the 4-center of mass of the universe ({\it decoupled point
particle clock}) reminding the Machian statement that only
relative motions are dynamically relevant.}. Since it is not known
how to build a global atlas of coordinate charts for the group
manifold of diffeomorphism groups, it is not known either how to
express the $\xi^r$'s, $\pi_{\phi}$ and the DO in terms of the
original ADM canonical variables\footnote{This should be compared
to the Yang-Mills theory in case of a trivial principal bundle,
where the corresponding variables are defined by a path integral
over the original canonical variables \cite{33a,29a,30a}.}.

\bigskip

\subsection{The Gauge Fixings and their Chrono-Geometrical Interpretation.}

The four gauge fixings to the secondary constraints, when written
in the quasi-Shanmugadhasan canonical basis, have the following
meaning:\hfill\break

 i) the three gauge fixings for the parameters $\xi^r$ of the
spatial passive diffeomorphisms generated by the super-momentum
constraints correspond to the choice of a system of 3-coordinates
on $\Sigma_{\tau}$\footnote{Since the diffeomorphism group has no
canonical identity, this gauge fixing has to be done in the
following way. We choose a 3-coordinate system by choosing a
parametrization of the six components ${}^3g_{rs}(\tau ,\vec
\sigma )$ of the 3-metric in terms of {\it only three} independent
functions. This amounts to fix the three functional degrees of
freedom associated with the diffeomorphism parameters $\xi^r(\tau
,\vec \sigma )$. For instance, a 3-orthogonal coordinate system is
identified by ${}^3g_{rs}(\tau ,\vec \sigma ) = 0$ for $r \not= s$
and ${}^3g_{rr} = \phi^2\,  exp(\sum_{\bar a = 1}^2 \gamma_{r\bar
a} r_{\bar a})$. Then, we impose the gauge fixing constraints
$\xi^r(\tau ,\vec \sigma ) - \sigma^r \approx 0$ as a way of
identifying this system of 3-coordinates with a conventional
origin of the diffeomorphism group manifold. }. The time constancy
of these gauge fixings generates the gauge fixings for the shift
functions $n_r$ (determination of gravito-magnetism) while the
time constancy of the latter leads to the fixation of the Dirac
multipliers $\lambda^{\vec n}_r$; \hfill\break

ii) the gauge fixing to the super-hamiltonian constraint
determines $\pi_{\phi}$: it is a fixation of the form of
$\Sigma_{\tau}$. It amounts to the choice of one particular 3+1
splitting of $M^4$ as well as to {\it the choice of a notion of
simultaneity}, namely of a convention for the synchronization of
all the clocks lying on $\Sigma_{\tau}$. Since the time constancy
of the gauge fixing on $\pi_{\phi}$ determines the gauge fixing
for the lapse function $n$ (and then of the Dirac multiplier
$\lambda_n$), it follows a connection with the choice of the
standard of local proper time (see below).

\bigskip

Finally the gauge fixings to the rest-frame conditions
(\ref{III5}) have the following meaning:\medskip

iii) they completely determine a {\it global non-inertial
space-time laboratory} associated to the embedding $z^{\mu}(\tau
,\vec \sigma ) = X^{\mu}(\tau ) + F^{\mu}(\tau ,\vec \sigma )$
describing the 3+1 splitting selected by i) and ii).
\bigskip

In conclusion, in a completely fixed Hamiltonian gauge {\it all
the gauge variables} $\xi^r$, $\pi_{\phi}$, $n$, $n_r$ {\it become
uniquely determined functions of the DO} $r_{\bar a}(\tau ,\vec
\sigma )$, $\pi_{\bar a}(\tau ,\vec \sigma )$, which at this stage
are four arbitrary fields. {\it Conversely}, this entails that,
after such a fixation of the gauge $G$, {\it the functional form
of the DO in terms of the original variables becomes
gauge-dependent}. At this point it is convenient to denote them as
$r^G_{\bar a}$, $\pi^G_{\bar a}$.

As a consequence, a representative of a {\it Hamiltonian
kinematical or off-shell gravitational field}, in a given gauge
equivalence class, is parametrized by $r_{\bar a}$, $\pi_{\bar a}$
and is an element of a {\it conformal gauge orbit} (it contains
all the 3-metrics in a conformal 3-geometry) spanned by the gauge
variables $\xi^r$, $\pi_{\phi}$, $n$, $n_r$. Therefore, according
to the gauge interpretation based on constraint theory, a {\it
Hamiltonian kinematical or off-shell gravitational field} is an
equivalence class of 4-metrics modulo the Hamiltonian group of
gauge transformations, which contains a well defined conformal
3-geometry. Clearly, this is a consequence of the different
invariance properties of the ADM and Hilbert actions, even if they
generate the same equations of motion.

\medskip

Moreover, also the (unknown) solution $\phi (\tau ,\vec \sigma )$
of the Lichnerowicz equation becomes a uniquely determined
functional of the DO, and this implies that all the geometrical
tensors like the 3-metric ${}^3g_{rs}(\tau ,\vec \sigma )$, the
extrinsic curvature ${}^3K_{rs}(\tau ,\vec \sigma )$ of the
simultaneity surfaces $\Sigma_{\tau}$ (determining their {\it
final actual form}, see below), and the 4-metric ${}^4g_{AB}(\tau
,\vec \sigma )$ {\it become uniquely determined functionals of the
DO only}.

This is true in particular for the {\it weak ADM energy} $E_{ADM}
= \int d^3\sigma\, {\cal E}_{ADM}(\tau ,\vec \sigma )$, since the
energy density ${\cal E}_{ADM}(\tau ,\vec \sigma )$ depends not
only on the DO but also on $\phi$ and on the gauge variables
$\xi^r$ and $\pi_{\phi}$ (this is how the non-tensorial nature of
the energy density in general relativity reveals itself in our
approach). In a fixed gauge we get $E_{ADM} = \int d^3\sigma\,
{\cal E}^G_{ADM}(\tau ,\vec \sigma )$ and this becomes the
functional that rules the Hamilton equations \cite{37a} for the DO
in the completely fixed gauge

 \beq
 {{\partial r^G_{\bar a}(\tau ,\vec \sigma )}\over {\partial \tau }}
  = \{ r^G_{\bar a}(\tau ,\vec \sigma ), E_{\mathrm{ADM}}\}^*,
\quad {{\partial \pi^G_{\bar a}(\tau ,\vec \sigma )}\over
{\partial \tau}} = \{\pi^G_{\bar a}(\tau ,\vec \sigma ),
E_{\mathrm{ADM}}\}^*,
 \label{III8}
  \eeq

\noindent where $E_{\mathrm{ADM}}$ is intended as the restriction
of the weak {\it ADM} energy to $\Gamma_4$ and where the
$\{\cdot,\cdot\}^*$ are Dirac Brackets. By using the inversion of
the first set of Eqs.(\ref{III8}) to get $\pi^G_{\bar a} =
\pi^G_{\bar a}[r^G_{\bar b}, {{\partial r^G_{\bar b}}\over
{\partial \tau}}]$, we arrive at the second order in time
equations ${{\partial2 r^G_{\bar a}(\tau ,\vec \sigma )}\over
{\partial \tau2}} = F^G_{\bar a}[r^G_{\bar b}(\tau ,\vec \sigma ),
{{\partial r^G_{\bar b}(\tau ,\vec \sigma )}\over {\partial
\tau}}, spatial\, gradients\, of\, r^G_{\bar b}(\tau ,\vec \sigma
)]$, where the $F^G_{\bar a}$'s are {\it effective forces} whose
functional form depends on the gauge $G$.\bigskip

Thus, once we have chosen any surface of the foliation as initial
Cauchy surface $\Sigma_{\tau_o}$ and assigned the initial data
$r_{\bar a}(\tau_o ,\vec \sigma )$, $\pi_{\bar a}(\tau_o ,\vec
\sigma )$ of the DO, we can calculate the solution of the
Einstein-Hamilton equations corresponding to these initial data.
Having found the solution in a completely fixed gauge, besides the
values of the DO throughout space-time we get the value of the
extrinsic curvature ${}^3K_{rs}(\tau ,\vec \sigma )$ of the
simultaneity surfaces $\Sigma_{\tau}$ as an {\it extra bonus}.
Therefore, on-shell, in the gauge G with given initial data for
the DO on $\Sigma^{(G)}_{\tau_o}$, the leaves
$\Sigma^{(G)}_{\tau}$ of the 3+1 splitting are dynamically
determined in the adapted radar 4-coordinates $(\tau ,\vec \sigma
)$ (which, as shown in II, also determine the point-events of
$M^4$ in the gauge $G$). As said in footnote 14, the knowledge of
the lapse and shift functions and of the extrinsic curvature in
the gauge $G$ allows to find the embedding $z^{\mu}_G(\tau ,\vec
\sigma )$ of the simultaneity leaves and the 4-metric ${}^4g_{G\,
\mu\nu}(x_G)$ in the gauge $G$ in an arbitrary 4-coordinate system
$x^{\mu}_G = z^{\mu}_G(\tau ,\vec \sigma )$ of $M^4$. If we redo
all the calculations in another complete Hamiltonian gauge $G_1$
with adapted radar 4-coordinates $(\tau_1, {\vec \sigma}_1)$, then
we can find the embedding $z^{\mu}_{G_1}(\tau_1, {\vec \sigma}_1)$
of the new simultaneity surfaces $\Sigma^{(G_1)}_{\tau_1}$ and the
4-metric ${}^4g_{G_1\, \mu\nu}(x_{G_1})$ in another arbitrary
4-coordinate system $x^{\mu}_{G_1} = z^{\mu}_{G_1}(\tau_1, {\vec
\sigma}_1)$. The new  initial data for the DO on the new Cauchy
surface $\Sigma^{(G_1)}_{\tau_{1\, o}}$, corresponding to the same
universe identified by the initial data on
$\Sigma^{(G)}_{\tau_o}$, have to be extracted from the
requirements $x^{\mu}_{G_1} = x^{\mu}_{G_1}(x_G)$ and
${}^4g_{G_1\, \mu\nu}(x_{G_1}) = {{\partial x^{\alpha}_G}\over
{\partial x^{\mu}_{G_1}}}\, {{\partial x^{\beta}_G}\over {\partial
x^{\nu}_{G_1}}}\, {}^4g_{G\, \alpha\beta}(x_G)$.
\medskip

{\it This circular setting brings the ADM procedure to its end by
determining the "universe", corresponding to the given initial
conditions for the DO in every gauge and including the associated
admissible dynamical definitions of simultaneity, distant clocks
synchronization and gravito-magnetism}.
\medskip

It is important to stress, therefore, that the {\it complete
determination of the chrono-geometry} clearly depends upon the
solution of Einstein-Hamilton equations of motion i.e., once the
Hamiltonian formalism is fixed by the gauge choices, {\it upon the
initial conditions for the DO}. This implies that the admissible
notions of distant simultaneity turn out to be {\it dynamically
determined} as said in the Introduction. However, as stressed
there, {\it within the Hamiltonian approach to metric gravity},
different admissible {\it conventions} about distant simultaneity
within the {\it same universe} are merely {\it gauge-related} {\it
conventions}, corresponding to different {\it complete gauge
options} in analogy to what happens in a non-dynamical way within
the framework of parametrized Minkowski theories\footnote{See
Ref.\cite{12a} for a discussion of this point in special
relativity and the gauge nature of the admissible notions of
simultaneity in parametrized Minkowski theories.}. The admissible
dynamical simultaneity notions in our class of space-times are
much less in number than the non-dynamical admissible simultaneity
notions in special relativity: as shown in Section VIII of the
second paper in Ref.\cite{23a}, if Minkowski space-time is thought
of as a special solution (with vanishing DO) of Einstein-Hamilton
equations, then its allowed 3+1 splittings must have 3-conformally
flat simultaneity 3-surfaces (due to the vanishing of the DO the
Cotton-York tensor vanishes), a restriction absent in special
relativity considered as an autonomous  theory.

We believe that this result throws an interesting new light even
on the old - and outdated - debate about the so-called {\it
conventionality}  of distant simultaneity in special relativity
showing the trading of {\it conventionality} with {\it gauge
freedom}. It is clear that the mechanism of the complete
Hamiltonian gauge based on the 3+1 splitting of space-time plays a
crucial role here.

Of course, it rests to be shown how the above dynamical
determination can be enforced {\it in practice} to synchronize
actual clocks, i.e., essentially, how to generalize to the gravity
case the formal structure of Einstein-Reichenbach's convention.
This discussion is given in all details in Ref.\cite{12a}.

\bigskip

\vfill\eject

\section{On the Physical Interpretation of Dirac Observables and
Gauge Variables: Tidal-like and Inertial-like Effects.}

Let us now discuss with a greater detail the physical meaning of
the Hamiltonian gauge variables and DO.

As shown in Section III, the 20 off-shell canonical variables of
the ADM Hamiltonian description are naturally subdivided into {\it
two sets} by the quasi-Shanmugadhasan transformation:
\bigskip

i) The first set contains {\it seven off-shell Abelian Hamiltonian
gauge variables} whose conjugate momenta are seven Abelianized
{\it first class constraints}. The eighth canonical pair comprises
the variable in which the super-hamiltonian constraint has to be
solved (the conformal factor of the 3-metric, $\phi =
({}^3g)^{1/12}$) and its conjugate momentum as the eighth gauge
variable. Precisely, the gauge variables are: $\xi^r$,
$\pi_{\phi}$ (primary\, gauge variables), $n$, $n_r$ (secondary
gauge variables). Note that a {\it primary} gauge variable has its
arbitrariness described by a Dirac multiplier, while a {\it
secondary} gauge variable inherits the arbitrariness of the Dirac
multipliers through the Hamilton equations.

\bigskip

ii) The second set contains the {\it off-shell gauge invariant
(non-local and in general non-tensor) DO}: $r_{\bar a}(\tau ,\vec
\sigma )$, $\pi_{\bar a}(\tau ,\vec \sigma )$, $\bar a = 1,2$.
They satisfy hyperbolic Hamilton equations.\vskip 0.5cm

Let us stress again that the above subdivision of canonical
variables in two sets is a peculiar outcome of the
quasi-Shanmugadhasan canonical transformation which has no simple
counterpart within the {\it Lagrangian viewpoint} at the level of
the Hilbert action and/or of Einstein's equations: at this level
the only clear statement is whether or not the curvature vanishes.
As anticipated in the Introduction, this subdivision amounts to an
extra piece of (non-local) information which should be added to
the traditional wisdom of the {\it equivalence principle}
asserting the local impossibility of distinguishing gravitational
from inertial effects. Indeed, we shall presently see that it
allows to distinguish and visualize which aspects of the local
physical effects on test matter contain a {\it genuine
gravitational} component and which aspects depend solely upon the
choice of the {\it global non-inertial space-time laboratory} with
the associated atlas of 4-coordinate systems in a topologically
trivial space-time: these latter effects could then be named {\it
inertial}, in analogy with what happens in the non-relativistic
Newtonian case in {\it global rigid} non-inertial reference
frames. Recall again that, when a complete choice of gauge is
made, the gauge variables as well as any tensorial quantity become
fixed uniquely by the gauge-fixing procedure to functions of DO in
that gauge.

One should be careful in discussing this subject because the very
definition of {\it inertial force} (and {\it gravitational} as
well) seems rather unnatural in general relativity. We can take
advantage, however, from the circumstance that the Hamiltonian
point of view leads naturally to a re-reading of geometrical
features in terms of the traditional concept of {\it force}.

First of all, recall that we are still considering here the case
of pure gravitational field without matter. It is then natural
first of all to characterize as genuine gravitational effects
those which are directly correlated to the {\it DO}. It is also
crucial to stress that such purely gravitational effects {\it are
absent in Newtonian gravity}, where there are no autonomous
gravitational fields, i.e., fields not generated by matter
sources. It seems therefore plausible to trace {\it inertial}
(much better than {\it fictitious}, in the relativistic case)
effects to a pure off-shell dependence on the {\it Hamiltonian
gauge variables}\footnote{By introducing dynamical matter the
Hamiltonian procedure leads to distinguish among {\it
action-at-a-distance, gravitational, and inertial} effects, with
consequent relevant implications upon concepts like gravitational
passive and active masses and, more generally, upon the problem of
the origin of inertia. See Ref.\cite{51a} for other attempts of
separating inertial from tidal effects in the equations of motion
in configuration space for test particles, in a framework in which
asymptotic inertial observers are refuted. In this reference one
finds also the following version (named Mach 11) of the Mach
principle: "The so-called {\it inertial effects}, occurring in a
non-inertial frame, are gravitational effects caused by the
distribution and motion of the distant matter in the universe,
relative to the frame". Thus {\it inertial} means here non-tidal +
true gravitational fields generated by cosmic matter. In the above
reference it is also suggested that super-fluid Helium II may be
an alternative to fixed stars as a standard of non rotation. Of
course all these interpretations are questionable. On the other
hand, the Hamiltonian framework offers the tools for making such a
distinction while distant matter effects are hidden in the
non-locality of DO and gauge variables. Since in a fixed gauge the
gauge variables are functions of the DO in that gauge, {\it tidal
effects are clearly mixed with inertial ones}. For a recent
critical discussion about the origin of inertia and its connection
with inertial effects in accelerated and rotating frames see
Ref.\cite{52a}.}. Recall also that, at the non-relativistic level,
Newtonian gravity is fully described by {\it action-at-a-distance
forces} and, in absence of matter, there are no {\it tidal} forces
among test particles. Tidal-like forces are {\it entirely}
determined by the variation of the action-at-a-distance force
created by the Newton potential of a massive body on the test
particles. In vacuum general relativity instead the {\it geodesic
deviation equation} shows that {\it tidal forces}, locally
described by the Riemann tensor, {\it act on test particles even
in absence of any kind of matter}.

\bigskip

Indeed fixing the off-shell Hamiltonian gauge variables determines
the weak ADM energy density ${\cal E}^G_{ADM}(\tau ,\vec \sigma )$
and the Hamilton equations (\ref{III8}). Therefore, from these
equations the form of the {\it effective inertial forces}
$F^G_{\bar a}$ is uniquely determined: they describe the {\it
form} in which physical gravitational effects determined by the DO
show themselves. Such {\it appearances} undergo {\it inertial}
changes upon going  from one global non-inertial reference frame
to another. Furthermore genuine gravitational effects are always
necessarily dressed by inertial-like appearances. Thus, the
situation is only vaguely analogous to the phenomenology of
non-relativistic inertial forces. These latter describe {\it
purely apparent} (or really {\it fictitious}) mechanical effects
which show up in accelerated Galilean reference frames
\footnote{With arbitrary {\it global} translational and rotational
3-accelerations.} and can be eliminated by going to (global)
inertial reference frames \footnote{See Ref.\cite{53a} for the
determination of {\it quasi-inertial reference frames} in
astronomy as those frames in which rotational and linear
acceleration effects lie under the sensibility threshold of the
measuring instruments.}. Besides the existence of autonomous
gravitational degrees of freedom, it is therefore clear that the
further deep difference concerning inertial-like forces in the
general-relativistic case with respect to Newtonian gravity rests
upon the fact that now inertial reference frames exist only
locally if freely falling along 4-geodesics.

\bigskip

For the sake of clarity, consider the non-relativistic Galilean
framework in greater detail. If a {\it global non-inertial
reference frame} has translational acceleration $\vec w(t)$ and
angular velocity $\vec \omega (t)$ with respect to a given
inertial frame, a particle with free motion ($\vec a = {\ddot
{\vec x}} = 0$) in the inertial frame has the following
acceleration as seen from the non-inertial frame

\beq
 {\vec a}_{NI} = - \vec w(t) + \vec x \times {\dot {\vec
\omega}}(t) + 2 {\dot {\vec x}} \times \vec \omega (t) + \vec
\omega (t) \times [\vec x \times \vec \omega (t)].
 \label{VI2}
 \eeq

\noindent After multiplication of this equation by the particle
mass, the second term on the right hand side is the {\it Jacobi
force}, the third is the {\it Coriolis force} and the fourth the
{\it centrifugal force}.

We have given in Ref.\cite{54a} a description of non-relativistic
gravity which is generally covariant under {\it arbitrary} passive
Galilean coordinate transformations [$t^{'} = T(t)$, ${\vec x}^{'}
= \vec f (t, \vec x)$]. The analogue of Eq.(\ref{VI2}) in this
case contains more general {\it apparent} forces, which are
reduced to those appearing in Eq.(\ref{VI2}) in particular rigid
coordinate systems. The discussion given in Ref.\cite{54a} is a
good introduction to the relativistic case, just because in
general relativity there are no {\it global inertial reference
frames}.
\bigskip

Two different approaches have been considered in the literature in
the general relativistic case concerning the choice of reference
frames, namely using either

\bigskip

i) a single accelerated time-like observer with an arbitrary
associated tetrad,

\vskip 0.5cm or \vskip 0.5cm

ii) a congruence of accelerated time-like observers with a
conventionally chosen associated field of tetrads\footnote{The
time-like tetrad field is the 4-velocity field of the congruence.
The conventional choice of the spatial triad is equivalent to a
choice of a specific system of gyroscopes (see footnote 45 in
Appendix A for the definition of a Fermi-Walker transported
triad). See the local interpretation in Ref.\cite{22a} of {\it
inertial} forces as effects depending on the {\it choice of a
congruence of time-like observers with their associated tetrad
fields} as a reference standard for their description. Note that,
in gravitational fields without matter, gravito-magnetic effects
as described by ${}^4g_{\tau r}$ are purely inertial effects in
our sense, since are determined by the {\it shift} gauge
variables. While in metric gravity the tetrad fields are used only
to rebuild the 4-metric, the complete theory taking into account
all the properties of the tetrad fields is {\it tetrad gravity}
\cite{23a}.}.

\bigskip

Usually, in both approaches the observers are {\it test
observers}, which describe phenomena from their kinematical point
of view without generating any dynamical effect on the system.
\bigskip

i) Consider first the case of a single {\it test observer} with
his tetrad (see Ref.\cite{55a,56a}).

\medskip

After the choice of the associated local Minkowskian system of
(Riemann-Gaussian) 4-coordinates, the line element
becomes\footnote{If the test observer is in free fall ({\it
geodesic observer}) we have $\vec a = 0$. If the triad of the test
observer is Fermi-Walker transported (standard of {\it
non-rotation} of the gyroscope) we have $\vec \omega = 0$. } $ds^2
= - \delta_{ij} dx^i dx^j + 2 \epsilon_{ijk} x^j {{\omega^k}\over
c} dx^o dx^i + [1 + {{2 \vec a \cdot \vec x}\over {c2}}
(dx^o)^2]$. The {\it test observer} describes a nearby time-like
geodesics $y^{\mu}(\lambda )$ ($\lambda$ is the affine parameter
or proper time) followed by a test particle in {\it free fall} in
a given gravitational field by means of the following spatial
equation: ${{d^2 \vec y}\over {(dy^o)^2}} = - \vec a - 2 \vec
\omega \times {{d \vec y}\over {dy^o}} + {2\over {c^2}} \Big( \vec
a \cdot {{d\vec y}\over {dy^o}}\Big)\, {{d\vec y}\over {dy^o}}$.
Thus, the relative acceleration of the particle with respect to
the observer with this special system of coordinates\footnote{It
replaces the global non-inertial non-relativistic reference frame.
With other coordinate systems, other terms would of course
appear.} is composed by the observer 3-acceleration plus a
relativistic correction and by a Coriolis acceleration\footnote{
This is caused by the rotation of the spatial triad carried by the
observer relative to a Fermi-Walker transported triad. The
vanishing of the Coriolis term justifies the statement that for an
observer which is not in free fall ($\vec a \not= 0$) a local
coordinate system produced by Fermi-Walker transport of the
spatial triad of vectors is the best possible realization of a
non-rotating system.}.  Note that, from the Hamiltonian point of
view, the constants $\vec a$ and $\vec \omega$ are constant
functionals of the DO of the gravitational field in this
particular gauge.

\medskip

As said above, different Hamiltonian gauge fixings on-shell,
corresponding to on-shell variations of the Hamiltonian gauge
variables, give rise to different appearances of the physical
effects as gauge-dependent functionals of the DO in that gauge of
the type ${\cal F}_{G}(r_{\bar a}, \pi_{\bar a})$ (like $\vec a$
and $\vec \omega$ in the previous example). \vskip 0.5truecm

In absence of matter, we can consider the {\it zero curvature
limit}, which is obtained by putting the DO to zero. In this way
we get Minkowski space-time (a solution of Einstein's equations)
equipped with those kinds of coordinates systems which are
compatible with Einstein's theory \footnote{As shown in
Ref.\cite{13a} this implies the vanishing of the Cotton-York
3-conformal tensor, namely the condition that the allowed 3+1
splittings of Minkowski space-time compatible with Einstein's
equations have the leaves {\it 3-conformally flat} in absence of
matter. This solution of Einstein's equations, has been named {\it
void space-time} in Ref.\cite{23a}: Minkowski space-time in
Cartesian 4-coordinates is just a gauge representative of it. Note
that, even if Einstein always rejected this concept, a void
space-time corresponds to the description of a {\it special class}
of 4-coordinate systems for Minkowski space-time without matter.
As a consequence  special relativity, considered as an autonomous
theory, admits much more general inertial effects associated with
the admissible 3+1 splittings of Minkowski space-time \cite{12a}
whose leaves are not 3-conformally flat.}. In particular, the
quantities ${\cal F}_G = lim_{r_{\bar a}, \pi_{\bar a} \rightarrow
0}\, {\cal F}_G(r_{\bar a}, \pi_{\bar a})$ describe inertial
effects in those 4-coordinate systems for Minkowski space-time
which have a counterpart in Einstein general relativity.

\medskip

In presence of matter Newtonian gravity is recovered with a double
limit:\medskip

a) the limit in which DO are restricted to the solutions of the
Hamilton equations (\ref{III7}) with matter, $r_{\bar a}
\rightarrow f_{\bar a}(matter)$, $\pi_{\bar a} \rightarrow g_{\bar
a}(matter)$, so that their insertion in the Hamilton equations for
matter produces effective gauge-dependent action-at-a-distance
forces;

b) the  $c \rightarrow \infty$ limit, in which curvature effects,
described by matter after the limit a), disappear, so that the
final action-at-a-distance forces are the Newtonian ones. \vskip
0.5truecm

This implies that the functionals ${\cal F}_G(r_{\bar a},
\pi_{\bar a})$ must be restricted to the limit ${\cal F}_{Newton}
= lim_{c \rightarrow \infty}\,\, lim_{r_{\bar a} \rightarrow
f_{\bar a}, \pi_{\bar a} \rightarrow g_{\bar a}}\,\,\Big( {\cal
F}_{G\,o} + {1\over c} {\cal F}_{G\, 1} + ...\Big) = {\cal F}_{G\,
o} {|}_{r_{\bar a} = f_{\bar a}, \pi_{\bar a} =  g_{\bar a}}$.
Then ${\cal F}_{Newton}$, which may be coordinate dependent,
becomes the {\it Newtonian inertial force} in the corresponding
general Galilean coordinate system.
\bigskip

ii) Consider then the more general case of a {\it congruence of
accelerated time-like observers} which is just the case with
reference to our {\it global non-inertial space-time laboratory}.
In this way it is possible to get a much more accurate and
elaborate description of the relative 3-acceleration, as seen in
his own local rest frame by each observer of the congruence which
intersects the geodesic of a test particle in free fall (see
Ref.\cite{22a}). The identification of various types of 3-forces
depends upon:\medskip

a) the gravitational field (the form of the geodesics obviously
depends on the metric tensor; usually the effects of the
gravitational field are classified as gravito-electric and
gravito-magnetic, even if this is strictly valid only in harmonic
coordinates),

\medskip
b) the properties (acceleration, vorticity, expansion, shear) of
the congruence of observers,

\medskip
c) the choice of the time-parameter used to describe the particle
3-trajectory in the local observer rest frame.
\bigskip

\noindent There are, therefore, many possibilities for defining
the relative 3-acceleration (see Ref.\cite{22a}) and its
separation in various types of inertial-like accelerations (See
Appendix A for a more complete discussion of the properties of the
congruences of time-like observers).
\bigskip

Summarizing, once a local reference frame has been chosen, in
every 4-coordinate system we can consider: \medskip

a) the {\it genuine tidal gravitational effects} which show up in
the geodesic deviation equation: they are well defined
gauge-dependent functionals of the DO associated to that gauge; DO
could then be called {\it non-local tidal-like degrees of
freedom};

b) the fact that {\it geodesic curves} will have different
geometrical descriptions corresponding to different gauges (i.e.
different inertial forces), although they will be again
functionally dependent only on the DO in the relevant gauge;

c) the issue of the description of the {\it relative
3-acceleration of a free particle in free fall}, as given in the
local rest frame of a generic observer of the congruence, which
will contain various terms. Such terms are identifiable with the
general relativistic extension of the various non-relativistic
kinds of inertial accelerations and all will again depend on the
DO in the chosen gauge, both directly and through the Hamiltonian
gauge variables of that gauge.
\bigskip

Three general remarks: \medskip

First of all, the picture we have presented is not altered by the
presence of matter. The only new phenomenon besides the above
purely gravitational, {\it inertial and tidal} effects, is that
from the solution of the super-hamiltonian and super-momentum
constraints emerge {\it action-at-a-distance, Newtonian-like and
gravito-magnetic} effects among matter elements, as already noted
in footnote 31.\medskip

Second, the {\it reference standards} of time and length
correspond to units of {\it coordinate time and length} and not to
proper times and proper lengths \cite{16a}: this is not in
contradiction with general covariance, because an extended {\it
laboratory}, in which one defines the reference standards,
corresponds to a particular {\it completely fixed on-shell
Hamiltonian gauge} plus a local congruence of time-like observers.
For instance, in astronomy and in the theory of satellites, the
unit of time is replaced by a unit of coordinate length ({\it
ephemerides time}). This leads to the necessity of taking into
account the theory of measurement in general relativity.

\medskip

Third, as evident from the structure of the Shanmugadhasan
transformation, the distinction between {\it tidal-like} and {\it
generalized non-inertial} effects is a gauge (i.e., a coordinate)
dependent concept. Although we deem this result to have physical
interest as it stands, the possibility remains open of pushing our
knowledge even further. Precisely, in the second paper we will
exploit to this effect a discussion about the relation between the
notion of DO and that of the so-called {\it Bergmann observables}
(BO)\cite{14a} which (although rather ambiguously) are defined to
be uniquely {\it predictable} from the initial data, but also
invariant under standard {\it passive diffeomorphisms} (PDIQ).

A possible starting point to attack the problem of the connection
of DO with BO seems to be a Hamiltonian re-formulation of the
Newman-Penrose formalism \cite{15a} (that contains only PDIQ)
employing Hamiltonian null-tetrads carried by the surface-forming
congruence of time-like observers. In view of this program, in
paper II we will argue in favor of a {\it main conjecture}
according to which special Darboux bases for canonical gravity
should exist in which the inertial effects (gauge variables) are
described by PDIQ while the autonomous degrees of freedom (DO) are
{\it also} BO. The hoped for validity of this conjecture would
amount - among other important consequences - to attributing to
our separation between {\it tidal-like} and {\it generalized
inertial effects} the status of an invariant statement. This would
give, in our opinion, a remarkable contribution to the long
standing debate about the equivalence principle.

\vfill\eject

\appendix

\section{Time-Like Accelerated Observers.}

In this Appendix we collect a number of scattered properties of
time-like observers.
\bigskip

An {\it inertial observer} in Minkowski space-time $M4$ is a
time-like future-oriented straight line $\gamma$ \cite{57a}. Any
point $P$ on $\gamma$ together with the unit time-like tangent
vector $e^{\mu}_{(o)}$ to $\gamma$ at $P$ is an {\it instantaneous
inertial observer}. Let us choose a point $P$ on $\gamma$ as the
origin of an {\it inertial system} $I_P$ having $\gamma$ as time
axis and three orthogonal space-like straight lines orthogonal to
$\gamma$ in $P$, with unit tangent vectors $e^{\mu}_{(r)}$,
$r=1,2,3$ as space axes. Let $x^{\mu}$ be a Cartesian 4-coordinate
system referred to these axes, in which the line element has the
form $ds2 = \eta_{\mu\nu}\, dx^{\mu}\, dx^{\nu}$ with
$\eta_{\mu\nu} = \epsilon\, (+---)$, $\epsilon = \pm 1$.
Associated to these coordinates there is a {\it reference frame}
(or {\it system of reference} or {\it platform} \cite{16a}) given
by the congruence of time-like straight lines parallel to
$\gamma$, namely a unit vector field $u^{\mu}(x)$. Each of the
integral lines of the vector field is identified by a fixed value
of the three spatial coordinates $x^i$ and represent an observer:
this is a {\it reference point} according to M$\o$ller \cite{58a}.
A reference frame $l$, i.e. a time-like vector field $l^{\mu}(x)\,
{{\partial}\over {\partial x^{\mu}}}$ with its  congruence of
time-like world-lines and its associated 1+3 splitting of $TM^4$,
admits the decomposition of Eq.(\ref{b3}) (see below).

 \medskip
 While in Newtonian physics an {\it absolute reference frame} is an
imagined extension of a rigid body and a clock (with any
coordinate systems attached), in general relativity \cite{59a} we
must replace the rigid body either by a cloud of test particles in
free fall ({\it geodesic congruence}) or by a test fluid ({\it
non-geodesic congruence} for non-vanishing pressure). Therefore a
{\it reference frame} is schematized as a future-pointing
time-like congruence with all the possible associated 4-coordinate
systems. This is called a {\it platform} in Ref.\cite{57a}, where
there is a classification of the possible types of platforms and
the definition of the position vector of a neighboring observer in
the local rest frame of a given observer of the platform. Then,
the Fermi-Walker covariant derivative (applied to a vector in the
rest frame it produces a new vector still in the rest frame
\cite{60a}) is used to define the 3-velocity (and then the
3-acceleration) of a neighboring observer in the rest frame of the
given observer, as the natural generalization of the Newtonian
relative 3-velocity (and 3-acceleration). See Ref.\cite{22a} for a
definition, based on these techniques, of the 3-acceleration of a
test particle in the local rest frame of an observer crossing the
particle geodesics, with the further introduction of the Lie and
co-rotating Fermi-Walker derivatives.

\bigskip

Consider now the point of view of the special (non-rotating,
surface-forming) congruence of time-like accelerated observers
whose 4-velocity field is the field of unit normals to the
space-like hyper-surfaces $\Sigma_{\tau}$.
\bigskip

We want to describe this non-rotating Hamiltonian congruence, by
emphasizing its interpretation in terms of gauge variables and DO.
The field of contravariant and covariant unit normals to the
space-like hyper-surfaces $\Sigma_{\tau}$ are expressed only in
terms of the lapse and shift gauge variables (as in Sections III
and IV, we use coordinates adapted to the foliation: $l^A(\tau
,\vec \sigma ) = b^A_{\mu}(\tau ,\vec \sigma )\, l^{\mu}(\tau
,\vec \sigma )$ with the $b^A_{\mu}(\tau ,\vec \sigma ) =
{{\partial \sigma^A}\over {\partial z^{\mu}}}$ being the
transition coefficients from adapted to general coordinates )

 \bea
 &&l^A(\tau ,\vec \sigma ) =  {1\over {N(\tau ,\vec \sigma
 )}}\, \Big( 1; -N^r(\tau ,\vec \sigma )\Big),\nonumber \\
 &&l_A(\tau ,\vec \sigma ) = N(\tau ,\vec \sigma )\, \Big( 1;
 0\Big),\qquad l^A(\tau ,\vec \sigma )\, l_A(\tau ,\vec \sigma ) =
 1.
 \label{b1}
 \eea

\medskip

Since this congruence is surface forming by construction, it has
zero vorticity and is non-rotating (in the sense of congruences).
As said in Section III, in Christodoulou-Klainermann space-times
\cite{9a} we have $N(\tau ,\vec \sigma ) = \epsilon + n(\tau ,\vec
\sigma )$, $N^r(\tau ,\vec \sigma ) = n^r(\tau ,\vec \sigma )$.
The specific time-like direction identified by the normal has
inertial-like nature, in the sense of being dependent on
Hamiltonian gauge variables only. Therefore the world-lines of the
observers of this foliation\footnote{ It is called the
Wigner-Sen-Witten (WSW) foliation\cite{13a} due to its properties
at spatial infinity (see footnote 14). The associated observers
are called {\it Eulerian observers} when a perfect fluid is
present as dynamical matter.} change on-shell going from a
4-coordinate system to another. On the other hand, the embeddings
$z^{\mu}_{\Sigma}(\tau ,\vec \sigma )$ of the leaves
$\Sigma_{\tau}$ of the WSW foliation in space-time depend on {\it
both} the DO and the gauge variables.
\medskip

If $x^{\mu}_{{\vec \sigma}_o}(\tau )$ is the time-like world-line
of the observer crossing the leave $\Sigma_{\tau_o}$ at ${\vec
\sigma}_o$, we have \footnote{ Note that the mathematical time
parameter $\tau$ labeling the leaves of the foliation is not in
general the proper time of any observer of the congruence.}

\bea
 x^{\mu}_{{\vec \sigma}_o}(\tau ) &=& z^{\mu}_{\Sigma}(\tau ,
 {\vec \rho}_{{\vec \sigma}_o}(\tau )),\quad with\,\, {\vec
 \rho}_{{\vec \sigma}_o}(\tau_o) = {\vec \sigma}_o,\qquad
 {\dot x}^{\mu}_{{\vec \sigma}_o}(\tau ) = {{d x^{\mu}_{{\vec
 \sigma}_o}(\tau )}\over {d \tau}},\nonumber \\
 &&{}\nonumber \\
 l^{\mu}_{{\vec \sigma}_o}(\tau ) &=& l^{\mu}(\tau , {\vec
 \rho}_{{\vec \sigma}_o}(\tau )) = {{{\dot x}^{\mu}_{{\vec
 \sigma}_o}(\tau )}\over {\sqrt{{}^4g_{\alpha\beta}(x_{{\vec \sigma}_o}(\tau ))\,
 {\dot x}^{\alpha}_{{\vec \sigma}_o}(\tau )\, {\dot x}^{\beta}_{{\vec \sigma}_o}(\tau
 )} }},\nonumber \\
 &&{}\nonumber \\
 &&a^{\mu}_{{\vec \sigma}_o}(\tau ) = {{d l^{\mu}_{{\vec
 \sigma}_o}(\tau )}\over {d \tau}},\qquad a^{\mu}_{{\vec
 \sigma}_o}(\tau )\, l_{{\vec \sigma}_o\, \mu}(\tau ) = 0.
 \label{b2}
 \eea

\noindent Here $a^{\mu}_{{\vec \sigma}_o}(\tau )$ is the
4-acceleration of the observer $x^{\mu}_{{\vec \sigma}_o}(\tau )$.
\medskip

As for any congruence, we have the decomposition ($P_{\mu\nu} =
\eta_{\mu\nu} - l_{\mu}\, l_{\nu}$)

\bea
 {}^4\nabla_{\mu}\, l_{\nu} &=&  l_{\mu}\, a_{\nu} + {1\over
3}\, \Theta \, P_{\mu\nu} + \sigma_{\mu\nu} +
\omega_{\mu\nu},\nonumber \\
 &&a^{\mu} = l^{\nu}\, {}^4\nabla_{\nu}\, l^{\mu} = {\dot l}^{\mu},
 \nonumber \\
 &&\Theta = {}^4\nabla_{\mu}\, l^{\mu},\nonumber \\
 &&\sigma_{\mu\nu} = {1\over 2}\, (a_{\mu}\, l_{\nu} + a_{\nu}\,
 l_{\mu}) + {1\over 2}\, ({}^4\nabla_{\mu}\,
l_{\nu} + {}^4\nabla_{\nu}\, l_{\mu}) - {1\over 3}\, \Theta\,
P_{\mu\nu},\nonumber \\
 && with\,\, magnitude\,\, \sigma^2={1\over 2} \sigma_{\mu\nu}\sigma^{\mu\nu},\nonumber \\
 &&\omega_{\mu\nu} = - \omega_{\nu\mu} = \epsilon_{\mu\nu\alpha\beta}\,
 \omega^{\alpha}\, l^{\beta} = {1\over 2}\, (a_{\mu}\, l_{\nu} - a_{\nu}\,
 l_{\mu}) + {1\over 2}\, ({}^4\nabla_{\mu}\, l_{\nu} - {}^4\nabla_{\nu}\,
  l_{\mu}) = 0,\nonumber \\
 && \omega^{\mu} = {1\over 2}\, \epsilon^{\mu\alpha\beta\gamma}\, \omega_{\alpha\beta}\,
l_{\gamma} = 0,
 \label{b3}
 \eea

\noindent where $a^{\mu}$ is the 4-acceleration, $\Theta$ the {\it
expansion} (it measures the average expansion of the
infinitesimally nearby world-lines surrounding a given world-line
in the congruence), $\sigma _{\mu\nu}$ the {\it shear} (it
measures how an initial sphere in the tangent space to the given
world-line, which is Lie transported along $l^{\mu}$ \footnote{It
has zero Lie derivative with respect to $l^{\mu}\,
\partial_{\mu} $.}, is distorted towards an ellipsoid with principal axes
given by the eigenvectors of $\sigma^{\mu}{}_{\nu}$, with rate
given by the eigenvalues of $\sigma^{\mu}{}_{\nu}$) and
$\omega_{\mu\nu}$ the {\it twist or vorticity} (it measures the
rotation of the nearby world-lines infinitesimally surrounding the
given one); $\sigma_{\mu\nu}$ and $\omega _{\mu\nu}$ are purely
spatial ($\sigma_{\mu\nu} l^{\nu} = \omega_{\mu\nu} l^{\nu} = 0$).
Due to the Frobenius theorem, the congruence is (locally)
hyper-surface orthogonal if and only if $\omega_{\mu\nu}=0$. The
equation ${1\over l}\, l^{\mu}\, \partial_{\mu}\, l = {1\over 3}\,
\Theta$ defines a representative length $l$ along the world-line
of $l^{\mu}$, describing the volume expansion (or contraction)
behaviour of the congruence.

While all these quantities depend on the Hamiltonian gauge
variables, the expansion and the shear depend a priori also upon
the DO, because  the covariant derivative is used in their
definition.

\bigskip

Yet, the ADM canonical formalism provides {\it additional
information}. Actually, on each space-like hyper-surface
$\Sigma_{\tau}$ of the foliation, there is a {\it privileged
contravariant space-like direction} identified by the lapse and
shift gauge variables \footnote{The unit vector ${\cal N}
^{\mu}(\tau ,\vec \sigma )$ contains a DO dependence in the
overall normalizing factor. The existence of this space-like gauge
direction seems to indicate that {\it synchronous or time
orthogonal} 4-coordinates with $N_r(\tau ,\vec \sigma ) = -
{}^4g_{\tau r}(\tau ,\vec \sigma ) = 0$ (absence of
gravito-magnetism) have singular nature \cite{61a}. Note that the
evolution vector of the {\it slicing point of view} has $N(\tau
,\vec \sigma )\, l^{\mu}(\tau ,\vec \sigma )$ and $|\vec N(\tau
,\vec \sigma )|\, {\cal N}^{\mu}(\tau ,\vec \sigma )$ as
projections along the normal and the plane tangent to
$\Sigma_{\tau}$, respectively.}

\bea
 {\cal N}^{\mu}(\tau ,\vec \sigma ) &=& {1\over {|\vec N(\tau
,\vec \sigma )|}}\, \Big( 0; n^r(\tau ,\vec \sigma
)\Big),\nonumber \\
 {\cal N}_{\mu}(\tau ,\vec \sigma ) &=& |\vec N(\tau ,\vec \sigma
 )|\, \Big( 1; {{N_r(\tau ,\vec \sigma )}\over {|\vec N(\tau ,\vec
 \sigma )|^2}}\Big),\nonumber \\
 &&{\cal N}^{\mu}(\tau ,\vec \sigma )\, l_{\mu}(\tau ,\vec \sigma
 ) = 0,\qquad {\cal N}^{\mu}(\tau ,\vec \sigma )\, {\cal N}_{\mu}(\tau ,\vec \sigma )
 = - 1,\nonumber \\
 && |\vec N(\tau ,\vec \sigma )| = \sqrt{({}^3g_{rs}\, N^r\, N^s)(\tau ,\vec \sigma
 )}.
 \label{b4}
 \eea
\medskip

If 4-coordinates, corresponding to an on-shell complete
Hamiltonian gauge fixing, exist such that the vector field defined
by ${\cal N}^{\mu}(\tau ,\vec \sigma )$ on each $\Sigma_{\tau}$ is
surface-forming (zero vorticity\footnote{ This requires that
${\cal N}_{\mu}\, dx^{\mu}$ is  a closed 1-form, namely that in
adapted coordinates we have $\partial_{\tau}\, {{N_r}\over {|\vec
N|}} = \partial_r\, |\vec N|$ and $\partial_r\, {{N_s}\over {|\vec
N|}} = \partial_s\, {{N_r}\over {|\vec N|}} $. This requires in
turn ${{N_r}\over {|\vec N|}} = \partial_r\, f$ with
$\partial_{\tau}\, f = |\vec N| + const.$}), then each
$\Sigma_{\tau}$ can be foliated with 2-surfaces, and the 3+1
splitting of space-time becomes a (2+1)+1 splitting corresponding
to the 2+2 splittings studied by Stachel and d'Inverno \cite{62a}.
\medskip

We have therefore a natural candidate for {\it one} of the three
spatial vectors of each observer, namely: $E^{\mu}_{{\vec
\sigma}_o\, ({\cal N})}(\tau ) = {\cal N}^{\mu}_{{\vec
\sigma}_o}(\tau ) = {\cal N}^{\mu}(\tau ,{\vec \rho}_{{\vec
\sigma}_o}(\tau ))$. By means of $l^{\mu}_{{\vec \sigma}_o}(\tau )
= l^{\mu}(\tau ,{\vec \rho}_{{\vec \sigma}_o}(\tau ))$ and ${\cal
N}^{\mu}_{{\vec \sigma}_o}(\tau )$, we can construct two {\it null
vectors} at each space-time point

\bea
 &&{\cal K}^{\mu}_{{\vec \sigma}_o}(\tau ) = \sqrt{{{|\vec N|}\over
 2}}\, \Big( l^{\mu}_{{\vec \sigma}_o}(\tau ) + {\cal N}^{\mu}_{{\vec
\sigma}_o}(\tau ) \Big),\nonumber \\
 &&{\cal L}^{\mu}_{{\vec \sigma}_o}(\tau ) = {1\over {\sqrt{2\, |\vec
 N|}}}\, \Big( l^{\mu}_{{\vec \sigma}_o}(\tau ) - {\cal N}^{\mu}_{{\vec
\sigma}_o}(\tau ) \Big).
 \label{b5}
 \eea

\noindent and then get a {\it null tetrad} of the type used in the
Newman-Penrose formalism \cite{15a}. The last two axes of the
spatial triad can be chosen as two space-like circular complex
polarization vectors $E^{\mu}_{{\vec \sigma}_o\, (\pm )}(\tau )$,
like in electromagnetism. They are built starting from the
transverse helicity polarization vectors $E^{\mu}_{{\vec
\sigma}_o\, (1,2)}(\tau )$, which are the first and second columns
of the standard Wigner helicity boost generating ${\cal
K}^{\mu}_{{\vec \sigma}_o}(\tau )$ from the reference vector
${\buildrel \circ \over {\cal K}}{}^{\mu}_{{\vec \sigma}_o}(\tau )
= |\vec N|\, \Big( 1; 001\Big)$ (see for instance the Appendices
of Ref.\cite{63a}).
\bigskip

Let us call $E^{(ADM) \mu}_{{\vec \sigma}_o\, (\alpha )}(\tau )$
the ADM tetrad formed by $l^{\mu}_{{\vec \sigma}_o}(\tau )$,
${\cal N}^{\mu}_{{\vec \sigma}_o}(\tau )$, $E^{\mu}_{{\vec
\sigma}_o\, (1,2)}(\tau )$ \footnote{It is a tetrad in adapted
coordinates: if $E^{\mu}_{(\alpha )} = {{\partial
z^{\mu}_{\Sigma}}\over {\partial \sigma^A}}\, E^A_{(\alpha )}$,
then $E^{(ADM)\, A}_{{\vec \sigma}_o\, (\alpha )}(\tau )\,$
${}^4g_{AB}(\tau , {\vec \rho}_{{\vec \sigma}_o}(\tau ))\,
E^{(ADM)\, B}_{{\vec \sigma}_o\, (\beta )}(\tau ) =
{}^4\eta_{(\alpha )(\beta )}$.}. This tetrad will not be in
general Fermi-Walker transported along the world-line
$x^{\mu}_{{\vec \sigma}_o}(\tau )$ of the observer\footnote{Given
the 4-velocity $l^{\mu}_{{\vec \sigma}_o}(\tau ) = E^{\mu}_{{\vec
\sigma}_o}(\tau )$ of the observer, the spatial triads
$E^{\mu}_{{\vec \sigma}_o\, (a)}(\tau )$, $a = 1,2,3$, have to be
chosen in a conventional way, namely by means of a conventional
assignment of an origin for the local measurements of rotations.
Usually, the choice corresponds to Fermi-Walker (FW) transported
({\it gyroscope-type transport, non-rotating observer}) tetrads
$E^{(FW)\, \mu}_{{\vec \sigma}_o\, (\alpha )}(\tau )$, such that
\break \hfill\break
\begin{eqnarray*}
{D\over {D\tau}}\, E^{(FW)\, \mu}_{{\vec \sigma}_o\, (a)}(\tau )
&=& \Omega^{(FW)}_{{\vec \sigma}_o}{}^{\mu}{}_{\nu}(\tau )\,
E^{(FW)\, \nu}_{{\vec \sigma}_o\, (a)}(\tau ) = l^{\mu}_{{\vec
\sigma}_o}(\tau )\, a_{{\vec \sigma}_o\, \nu}(\tau )\, E^{(FW)\,
\nu}_{{\vec \sigma}_o\, (a)}(\tau ),\nonumber \\
 &&\Omega^{(FW)}_{{\vec \sigma}_o}{}^{\mu\nu}(\tau ) =
 a^{\mu}_{{\vec \sigma}_o}(\tau )\, l^{\nu}_{{\vec \sigma}_o}(\tau
 ) - a^{\nu}_{{\vec \sigma}_o}(\tau )\, l^{\mu}_{{\vec \sigma}_o}(\tau
 ).
 \end{eqnarray*}
\hfill\break
 \noindent The triad $E^{(FW)\, \mu}_{{\vec
\sigma}_o\, (a)}(\tau )$ is the correct relativistic
generalization of {\it global Galilean non-rotating frames} (see
Ref.\cite{54a}) and is defined using only local geometrical and
group-theoretical concepts. Any other choice of the triads (Lie
transport, co-rotating-FW transport,...) is obviously also
possible \cite{22a}. A generic triad $E^{\mu}_{{\vec \sigma}_o\,
(a)}(\tau )$ will satisfy ${D\over {D\tau}}\, E^{\mu}_{{\vec
\sigma}_o\, (a)}(\tau ) = \Omega_{{\vec
\sigma}_o}{}^{\mu}{}_{\nu}(\tau )\, E^{\nu}_{{\vec \sigma}_o\,
(a)}(\tau )$ with $\Omega^{\mu\nu}_{{\vec \sigma}_o} =
\Omega^{(FW)\, \mu\nu}_{{\vec \sigma}_o} + \Omega^{(SR)\,
\mu\nu}_{{\vec \sigma}_o}$ with the spatial rotation part
$\Omega^{(SR)\, \mu\nu}_{{\vec \sigma}_o} =
\epsilon^{\mu\nu\alpha\beta}\, l_{{\vec \sigma}_o\, \alpha}\,
J_{{\vec \sigma}_o\, \beta}$, $J^{\mu}_{{\vec \sigma}_o}\,
l_{{\vec \sigma}_o\, \mu} = 0$, producing a rotation of the
gyroscope in the local space-like 2-plane orthogonal to
$l^{\mu}_{{\vec \sigma}_o}$ and $J^{\mu}_{{\vec \sigma}_o}$.}.
\medskip

Another possible (but only on-shell) choice of the spatial triad
together with the unit normal to $\Sigma_{\tau}$ is the {\it local
WSW (on-shell) compass of inertia} quoted in footnote 14, namely
the triads transported with the Frauendiener-Sen-Witten transport
(see footnote 73 and Eq.(12.2) of Ref.\cite{13a}) starting from an
asymptotic conventional triad (choice of the fixed stars) added to
the ADM 4-momentum at spatial infinity. As shown in Eq.(12.3) of
Ref.\cite{13a}, they have the expression $E^{(WSW) \mu}_{{\vec
\sigma}_o (a)}(\tau) = {{\partial z^{\mu}_{\Sigma}}\over {\partial
\sigma^s}} {|}_{x_{{\vec \sigma}_o}(\tau )}\, {}^3e^{(WSW)
s}_{{\vec \sigma}_o\, (a)}(\tau )$ where the triad
${}^3e^{(WSW)}_{{\vec \sigma}_o (a)}$ is solution of the
Frauendiener-Sen-Witten equation restricted to a solution of
Einstein equations.

\bigskip

Given an observer with world-line $x^{\mu}_{{\vec \sigma}_o}(\tau
)$ and tetrad $E^{\mu}_{{\vec \sigma}_o\, (\alpha )}(\tau )$, the
geometrical properties are described by the Frenet-Serret
equations \cite{64a}

\bea
 &&{D\over {D\tau}}\, l^{\mu}_{{\vec \sigma}_o}(\tau ) =
 \kappa_{{\vec \sigma}_o}(\tau )\, E^{\mu}_{{\vec \sigma}_o\,
 (1)}(\tau ),\nonumber \\
 &&{D\over {D\tau}}\,  E^{\mu}_{{\vec \sigma}_o\, (1)}(\tau ) =
 a^{\mu}_{{\vec \sigma}_o}(\tau ) =
 \kappa_{{\vec \sigma}_o}(\tau )\, l^{\mu}_{{\vec \sigma}_o}(\tau
 ) + \tau_{{\vec \sigma}_o\, (1)}(\tau )\,  E^{\mu}_{{\vec \sigma}_o\,
 (2)}(\tau ),\nonumber \\
 &&{D\over {D\tau}}\,  E^{\mu}_{{\vec \sigma}_o\, (2)}(\tau ) = -
 \tau_{{\vec \sigma}_o\, (1)}(\tau )\,  E^{\mu}_{{\vec \sigma}_o\,
 (1)}(\tau ) + \tau_{{\vec \sigma}_o\, (2)}(\tau )\,  E^{\mu}_{{\vec \sigma}_o\,
 (3)}(\tau ),\nonumber \\
 &&{D\over {D\tau}}\,  E^{\mu}_{{\vec \sigma}_o\, (3)}(\tau ) = -
 \tau_{{\vec \sigma}_o\, (2)}(\tau )\,  E^{\mu}_{{\vec \sigma}_o\,
 (2)}(\tau ),
 \label{b6}
 \eea

\noindent where $\kappa_{{\vec \sigma}_o}(\tau )$, $\tau_{{\vec
\sigma}_o (a)}(\tau )$, $a = 1,2$, are the curvature and the first
and second torsion of the world-line. $ E^{\mu}_{{\vec \sigma}_o\,
(a)}(\tau )$, $a = 1,2,3$ are said the normal and the first and
second bi-normal of the world-line, respectively.
\bigskip

Let us now look at the description of a geodesics $y^{\mu}(\tau
)$, the world-line of a scalar test particle, from the point of
view of those observers $\gamma_{{\vec \sigma}_o, y(\tau )}$ of
the congruence who intersect it, namely such that at $\tau$ it
holds $x^{\mu}_{{\vec \sigma}_o, y(\tau )}(\tau ) = y^{\mu}(\tau
)$. The family of these observers is called a {\it relative
observer world 2-sheet} in Ref.\cite{22a}.
\medskip

Since the parameter $\tau$ labeling the leaves $\Sigma_{\tau}$ of
the foliation is not the proper time $s = s(\tau )$ of the
geodesics $y^{\mu}(\tau ) = Y^{\mu}(s(\tau ))$, the geodesics
equation ${{d^2 Y^{\mu}(s)}\over {ds^2}} +
{}^4\Gamma^{\mu}_{\alpha\beta}(Y(s))\, {{dY^{\alpha}(s)}\over
{ds}}\, {{d Y^{\beta}(s)}\over {ds}} = 0$ (or $m\, a^{\mu}(s) =
m\, {{d^2 Y^{\mu}(s)}\over {ds^2}} = F^{\mu}(s)$, where $m$ is the
mass of the test particle), becomes

\bea
 &&{{d^2 y^{\mu}(\tau )}\over {d\tau^2}} +
{}^4\Gamma^{\mu}_{\alpha\beta}(y(\tau ))\,
{{dy^{\alpha}(\tau)}\over {d\tau}}\, {{d y^{\beta}(\tau )}\over
{d\tau}} - {{dy^{\mu}(\tau )}\over {d\tau}}\, {{d2 s(\tau
)}\over{d\tau2}}\, \Big( {{d s(\tau )}\over {d\tau}}\Big)^{-1} =
0,\nonumber \\
 &&{}
 \label{b7}
  \eea

or

\bea
 && m\, a_y^{\mu}(\tau ) = m\, {{d^2 y^{\mu}(\tau )}\over {d\tau^2}}
 =  f^{\mu}(\tau ).
 \label{b8}
 \eea

\noindent We see that the force $f^{\mu}(\tau )$ contains an
extra-piece with respect to $F^{\mu}(s(\tau ))$, due to the change
of time parameter. \vskip 0.5truecm

Let $U^{\mu}(\tau ) = V^{\mu}(s(\tau )) = {{d Y^{\mu}(s)}\over
{ds}} {|}_{s = s(\tau )} = {{{\dot y}^{\mu}(\tau)}\over
{\sqrt{{}^4g_{\alpha\beta}(y(\tau ))\, {\dot y}^{\alpha}(\tau )\,
{\dot y}^{\beta}(\tau ) }}}$ with ${\dot y}^{\mu}(\tau ) =
{{dy^{\mu}(\tau )}\over {d\tau}}$ be the 4-velocity of the test
particle and $ds = \sqrt{{}^4g_{\alpha\beta}(y(\tau ))\, {\dot
y}^{\alpha}(\tau )\, {\dot y}^{\beta}(\tau ) }\, d\tau$ be the
relation between the two parameters. By using the {\it intrinsic
or absolute derivative} along the geodesics parametrized with the
proper time $s = s(\tau )$ , the geodesics equation becomes ${\cal
A}^{\mu}(s) = {{D V^{\mu}(s)}\over {ds}} = 0$ [or ${\tilde {\cal
A}}^{\mu}(\tau ) = {{D U^{\mu}(\tau )}\over {d\tau}} =
{{dy^{\mu}(\tau )}\over {d\tau}}\, {{d2 s(\tau )}\over{d\tau2}}\,
\Big( {{d s(\tau )}\over {d\tau}}\Big)^{-1} = g^{\mu}(\tau ) $].

\medskip

In non-relativistic physics {\it spatial inertial forces are
defined as minus the spatial relative accelerations}, with respect
to an accelerated {\it global Galilean frame} (see
Ref.\cite{54a}). In general relativity one needs the whole
relative observer world 2-sheet to define an abstract 3-path in
the quotient space of space-time by the observer-family
world-lines, representing the trajectory of the test particle in
the observer {\it 3-space}. Moreover, a well defined projected
{\it time derivative} is needed to define a relative acceleration
associated to such 3-path. At each point $P(\tau )$ of the
geodesics, identified by a value of $\tau$, we have the two
vectors $U^{\mu}(\tau )$ and $l^{\mu}_{{\vec \sigma}_o\, y(\tau
)}(\tau )$. Therefore, each vector $X^{\mu}$ in the tangent space
to space-time in that point $P(\tau )$ admits two splittings:
\bigskip

i) $X^{\mu} = X_U\, U^{\mu} + P(U)^{\mu}{}_{\nu}\, X^{\nu}$,
$P^{\mu\nu}(U) = {}^4g^{\mu\nu} - U^{\mu}\, U^{\nu}$, i.e., into a
temporal component along $U^{\mu}$ and a spatial transverse
component, living in the local rest frame $LRS_U$;

ii) $X^{\mu} = X_l\, l^{\mu}_{{\vec \sigma}_o y(\tau )} +
P(l_{{\vec \sigma}_o y(\tau )})^{\mu}{}_{\nu}\, X^{\nu}$, i.e.,
into a temporal component along $l^{\mu}_{{\vec \sigma}_o y(\tau
)}(\tau )$ and a spatial transverse component, living in the local
rest frame $LRS_l$, which is the plane tangent to the leave
$\Sigma_{\tau}$ in $P(\tau )$ for our surface-forming congruence.
\bigskip

The {\it measurement} of $X^{\mu}$ by the observer congruence
consists in determining the scalar $X_l$ and the spatial
transverse vector. In adapted coordinates and after a choice of
the spatial triads, the spatial transverse vector is described by
the three (coordinate independent) tetradic components $X_{(a)} =
E^{\mu}_{(a)}\, X_{\mu}$. The same holds for every tensor.
Moreover, every spatial vector like $ P(U)^{\mu}{}_{\nu}\,
X^{\nu}$ in $LRS_U$ admits a 2+1 orthogonal decomposition ({\it
relative motion orthogonal decomposition}) into a component in the
2-dimensional rest subspace $LRS_U \cap LRS_L$ transverse to the
direction of relative motion and one component in the
1-dimensional (longitudinal) orthogonal complement along the
direction of the relative motion in each such rest space.
\medskip

At each point $P(\tau)$, the tangent space is split into the {\it
relative observer 2-plane} spanned by $U^{\mu}(\tau )$ and
$l^{\mu}_{{\vec \sigma}_o y(\tau )}(\tau )$ and into an orthogonal
space-like 2-plane. We have the 1+3 orthogonal decomposition

\bea
 U^{\mu}(\tau ) &=& \gamma (U,l)(\tau )\, \Big( l^{\mu}_{{\vec
 \sigma}_o\, y(\tau )}(\tau ) + \nu^{\mu}(U,l)(\tau
 )\Big),\nonumber \\
 &&\gamma (U,l) = U_{\mu}\, l^{\mu}_{{\vec \sigma}_o y(\tau
 )},\qquad \nu(U,l) = \sqrt{\nu^{\mu}(U,l)\, \nu_{\mu}(U,l)},\nonumber \\
 &&{\hat \nu}^{\mu}(U,l) = {{\nu^{\mu}(U,l)}\over {\nu
 (U,l)}},\qquad relative\,\,\, 4-velocity\,\,\, tangent\, to\, \Sigma_{\tau}.
 \label{b9}
 \eea

\bigskip

The equation of geodesics, written as $m\, {\cal A}^{\mu}(s) = 0$,
is described by the observers' family as:

i) a temporal projection along $l^{\mu}_{{\vec \sigma}_o y(\tau
)}$, leading to the evolution equation $m\, {\cal A}_{\mu}\,
l^{\mu}_{{\vec \sigma}_o y(\tau )} = 0$, for the observed energy
($E(U,l) = \gamma (U,l)$) of the test particle along its
world-line;

ii) a spatial projection orthogonal to $l^{\mu}_{{\vec \sigma}_o
y(\tau )}$ (tangent to $\Sigma_{\tau}$), leading to the evolution
equation for the observed 3-momentum of the test particle along
its world-line, with the kinematic quantities describing the
motion of the family of observers entering as {\it inertial
forces}. If, instead of writing $m\, P(l)^{\mu}{}_{\nu}\, {\cal
A}^{\mu}(s) = 0$ with $P(l)^{\mu\nu} = {}^4g^{\mu\nu} -
l^{\mu}_{{\vec \sigma}_o y(\tau )}\, l^{\nu}_{{\vec \sigma}_o
y(\tau )}$, we re-scale the particle proper time $s(\tau )$ to the
sequence of observer proper times $s_{(U,l)}$ defined by ${{d
s_{(U,l)}}\over {ds}} = \gamma (U,l)$, the spatial projection of
the geodesics equation, re-scaled with the gamma factor
\footnote{Namely $m\, \gamma^{-1}(U,l)\, P(l)^{\mu}{}_{\nu}\,
{\cal A}^{\nu} = 0$.}, can be written in the form

\bea
 m\, \Big( {{D_{(FW)}(U,l)\,}\over {d s_{(U,l)}}}\Big)^{\mu}{}_{\nu}\,
 v^{\nu}(U,l) &=& m\, a^{\mu}_{(FW)}(U,l) = F_{(FW)}^{(G)\mu}(U,l),\nonumber \\
 &&{}\nonumber \\
 F^{(G)\mu}_{(FW)}(U,l) &=& - \gamma (U,l)^{-1}\,
 P^{\mu}{}_{\nu}(l)\, {{D l^{\nu}_{{\vec \sigma}_o y(\tau )}(\tau (s))}\over
 {ds}} =\nonumber \\
 &=& - \Big( {{D_{(FW)}(U,l)}\over {ds_{(U,l)}}}\Big)^{\mu}{}_{\nu} \,
 l^{\nu}_{{\vec \sigma}_o y(\tau )}(\tau (s_{(U,l)})) =\nonumber \\
 &=& \gamma (U,l)\, \Big[ - a^{\mu}(l) + \Big( -
 \omega^{\mu}{}_{\nu}(l) + \theta^{\mu}{}_{\nu}(l) \Big)
 \nu^{\nu}(U,l) \Big],
 \label{b10}
 \eea

\noindent where $v^{\mu}(U,l) = U^{\mu} - \gamma (U,l)\,
l^{\mu}_{{\vec \sigma}_o y(\tau )} = v(U,l)\, {\hat
\nu}^{\mu}(U,l)$ with $v(U,l) = \gamma (U,l)\, \nu (U,l)$, and
$P(l)^{\mu}{}_{\nu}\, {D\over {ds}} = \Big( {{D_{(FW)}(U,l)}\over
{ds}}\Big)^{\mu}{}_{\nu}$ is the spatial FW intrinsic derivative
along the test world-line and $a^{\mu}_{(FW)}(U,l)$ is the {\it FW
relative acceleration}. The term $F^{(G)\mu}_{(FW)}(U,l)$ can be
interpreted as the set of {\it inertial forces} due to the motion
of the observers themselves, as in the non-relativistic case. Such
inertial forces depend on the following congruence properties:
\bigskip

i) the acceleration vector field $a^{\mu}(l)$, leading to a {\it
gravito-electric field and a spatial gravito-electric
gravitational force};

ii) the vorticity $\omega^{\mu}{}_{\nu}(l)$ and expansion + shear
$\theta^{\mu}{}_{\nu}(l) = \sigma^{\mu}_{\nu}(l) + {1\over 3}\,
\Theta (l)\, P^{\mu}{}_{\nu}(l)$ mixed tensor fields, leading to a
{\it gravito-magnetic vector field and tensor field and a Coriolis
or gravito-magnetic force} linear in the relative velocity
$\nu^{\mu}(U,l)$. \vskip 0.5truecm

Then, by writing $v^{\mu}(U,l) = v(U,l)\, {\hat \nu}^{\mu}(U,l)$,
the FW relative acceleration can be decomposed into a longitudinal
and a transverse relative acceleration

\bea
 a^{\mu}_{(FW)}(U,l) &=& {{D_{(FW)}(U,l)\, v(U,l)}\over
 {ds_{(U,l)}}}\, {\hat \nu}^{\mu}(U,l) + \gamma (U,l)\,
 a^{(\perp)\mu}_{(FW)}(U,l),\nonumber \\
 &&{}\nonumber \\
 a^{(\perp)\mu}_{(FW)}(U,l) &=& v(U,l)\, \Big({{D_{(FW)}}\over
 {ds_{(U,l)}}}\Big)^{\mu}{}_{\nu}\, {\hat \nu}^{\nu}(U,l)
 =\nonumber \\
 &=& \nu^2(U,l)\, \Big( {{D_{(FW)}}\over
 {dr_{(U,l)}}}\Big)^{\mu}{}_{\nu}\, {\hat \nu}^{\nu}(U,l) =
 {{\nu^2(U,l)}\over {\rho_{(FW)}(U,l)}}\, {\hat
 \eta}^{\mu}_{(FW)}(U,l).
 \label{b11}
 \eea

\noindent In the second expression of the transverse FW relative
acceleration, the reparametrization ${{dr_{(U,l)}}\over
{ds_{(U,l)}}} = \nu (U,L)$ to a spatial arc-length parameter has
been done. Since $\gamma (U,l)\, a^{(\perp)\mu}_{(FW)}(U,l)$ is
the transverse part of the relative acceleration, i.e. the {\it FW
relative centripetal acceleration}, $- m\, \gamma (U,l)\,
a^{(\perp)\mu}_{(FW)}(U,l)$ may be interpreted as a {\it
centrifugal force}, so that the geodesics equation is rewritten as
$m\, {{D_{(FW)}(U,l)\, v(U,l)}\over {ds_{(U,l)}}}\, {\hat
\nu}^{\mu}(U,l) = F^{(G)\mu}_{(FW)}(U,l) - m\, \gamma (U,l)\,
a^{(\perp)\mu}_{(FW)}(U,l)$, with the first member called
sometimes {\it Euler force}.
\medskip

The 3-path in the abstract quotient space can be treated as an
ordinary 3-curve in a 3-dimensional Riemann space. Its tangent is
${\hat \nu}^{\mu}(U,l)$, while its normal and bi-normal are
denoted ${\hat \eta}^{\mu}_{(FW)}(U,l)$ and ${\hat
\xi}^{\mu}_{(FW)}(U,l)$ respectively. The 3-dimensional
Frenet-Serret equations are then

\bea
 \Big( {{D_{(FW)}(U,l)}\over {dr_{(U,l)}}}\Big)^{\mu}{}_{\nu}\,
 {\hat \nu}^{\nu}(U,l) &=& \kappa_{(FW)}(U,l)\, {\hat
 \eta}^{\mu}_{(FW)}(U,l),\nonumber \\
 \Big( {{D_{(FW)}(U,l)}\over {dr_{(U,l)}}}\Big)^{\mu}{}_{\nu}\,
 {\hat \eta}^{\nu}_{(FW)}(U,l) &=& - \kappa_{(FW)}(U,l)\, {\hat
 \nu}^{\mu}(U,l) + \tau_{(FW)}(U,l)\, {\hat
 \xi}^{\mu}_{(FW)}(U,l),\nonumber \\
  \Big( {{D_{(FW)}(U,l)}\over {dr_{(U,l)}}}\Big)^{\mu}{}_{\nu}\,
 {\hat \xi}^{\nu}_{(FW)}(U,l) &=& - \tau_{(FW)}(U,l)\, {\hat
 \eta}_{(FW)}^{\mu}(U,l),
 \label{b12}
 \eea

\noindent where $\kappa_{(FW)}(U,l) = 1/ \rho_{(FW)}(U,l)$ and
$\tau_{(FW)}(U,l)$ are the curvature and torsion of the 3-curve,
respectively.

\bigskip

The main drawback of the 1+3 ({\it threading}) description,
notwithstanding its naturalness from a locally operational point
of view, is the use of a rotating congruence of time-like
observers: this introduces an element of non-integrability and, as
yet, no formulation of the Cauchy problem for the 1+3
re-formulation of Einstein's equations has been worked out.

\vfill\eject


\end{document}